\documentclass[12pt,letterpaper]{article}
\usepackage[includeheadfoot,  
            marginratio={1:1,2:3},  
            width=412pt,  
            height=688pt,]{geometry} 
\usepackage{amsmath} 
\usepackage{amsfonts} 
\usepackage{amssymb} 
\usepackage{bbm} 
 
\usepackage{graphicx} 
 
\allowdisplaybreaks[2] 
\numberwithin{equation}{section}

\newcommand{\ov}{\overline}

\newcommand{\thba}[2]{[\!\!\begin{array}{c} 
  {\phantom{}\vspace{-.5mm}\scriptstyle#1}% 
   \\[-1.6mm]{\scriptstyle #2}\end{array}\!\!]}

\newcommand{\be}{\begin{equation}} 
\newcommand{\ee}{\end{equation}} 
\newcommand{\bea}{\begin{eqnarray}} 
\newcommand{\eea}{\end{eqnarray}} 
\newcommand{\mbb}{\mathbb}

\def\IR{\relax{\rm I\kern-.18em R}} 
 
\def\ov#1{\overline{#1}} 
\def\IP{\relax{\rm I\kern-.18em P}} 
\def\inbar{\vrule height1.5ex width.4pt depth0pt} 
\def\IC{\relax\,\hbox{$\inbar\kern-.3em{\rm C}$}}

\def\K3{{\bf K3}}

\def\ov{\overline}

\DeclareMathOperator{\sign}{sign} 
 
\begin{document}

%%%%%%%%%%%%%%%%%%%%%%%%%%%%%%%%%%%%%%%%%%%%%%% 
%%%%%%%%%%%%%%%%%%%%%%%%%%%%%%%%%%%%%%%%%%%%%%% 
%%%%%%%%%%%%%%%%%%%%%%%%%%%%%%%%%%%%%%%%%%%%%%% 
%%%%%%%%%%%%%%%%%%%%%%%%%%%%%%%%%%%%%%%%%%%%%%% 
%%%%%%%%%%%%%%%%%%%%%%%%%%%%%%%%%%%%%%%%%%%%%%% 
%%%%%%%%%%%%%%%%%%%%%%%%%%%%%%%%%%%%%%%%%%%%%%% 
%%%%%%%%%%%%%%%%%%%%%%%%%%%%%%%%%%%%%%%%%%%%%%% 
%%%%%%%%%%%%%%%%%%%%%%%%%%%%%%%%%%%%%%%%%%%%%%% 

\vspace*{-1.5cm} 
\begin{flushright} 
  {\small 
  MPP-2007-57 \\ 
  LMU-ASC 31/07 \\ 
  } 
\end{flushright}

\vspace{1.5cm} 
\begin{center} 
  {\LARGE 
  Instantons and  Holomorphic Couplings in 
 Intersecting D-brane Models\\ 
  } 
\end{center}

\vspace{0.25cm} 
\begin{center} 
  {\small 
  Nikolas Akerblom$^1$, Ralph~Blumenhagen$^1$, Dieter L\"ust$^{1,2}$, \\ 
  Maximilian Schmidt-Sommerfeld$^1$ \\ 
  } 
\end{center}

\vspace{0.1cm} 
\begin{center} 
  \emph{$^{1}$ 
  Max-Planck-Institut f\"ur Physik, F\"ohringer Ring 6, \\ 
  80805 M\"unchen, Germany } \\ 
  \vspace{0.25cm} 
  \emph{$^{2}$ Arnold-Sommerfeld-Center for Theoretical Physics, \\ 
  Department f\"ur Physik, Ludwig-Maximilians-Universit\"at  M\"unchen, \\ 
  Theresienstra\ss e 37, 80333 M\"unchen, Germany} \\ 
\end{center}

\vspace{-0.1cm} 
\begin{center} 
  \tt{ 
  akerblom, blumenha, luest, pumuckl@mppmu.mpg.de \\ 
  } 
\end{center}

\vspace{1.5cm} 
\begin{abstract} 
\noindent  
We clarify certain aspects and discuss extensions of  the  
recently introduced string D-instanton calculus (hep-th/0609191). 
The one-loop determinants are related to  
one-loop open string threshold corrections in  
intersecting D6-brane models. 
Utilising  a non-renormalisation theorem 
for the  holomorphic Wilsonian gauge kinetic functions, 
we derive a number of constraints for the moduli dependence 
of the matter field K\"ahler potentials of intersecting D6-brane models on the 
torus. Moreover, we compute string one-loop corrections 
to the Fayet-Iliopoulos terms on the D6-branes 
finding that they are proportional to the  
gauge threshold corrections. 
 Employing these results, we discuss the issue of holomorphy 
for E2-instanton corrections to the superpotential.  
Eventually, we discuss E2-instanton corrections 
to the gauge kinetic functions and the FI-terms. 
\end{abstract}

\thispagestyle{empty} 
\clearpage 
\tableofcontents

%%%%%%%%%%%%%%%%%%%%%%%%%%%%%%%%%%%%%%%%%%%%%%% 
%%%%%%%%%%%%%%%%%%%%%%%%%%%%%%%%%%%%%%%%%%%%%%% 
%%%%%%%%%%%%%%%%%%%%%%%%%%%%%%%%%%%%%%%%%%%%%%% 
%%%%%%%%%%%%%%%%%%%%%%%%%%%%%%%%%%%%%%%%%%%%%%% 
%%%%%%%%%%%%%%%%%%%%%%%%%%%%%%%%%%%%%%%%%%%%%%% 
%%%%%%%%%%%%%%%%%%%%%%%%%%%%%%%%%%%%%%%%%%%%%%% 
%%%%%%%%%%%%%%%%%%%%%%%%%%%%%%%%%%%%%%%%%%%%%%% 
%%%%%%%%%%%%%%%%%%%%%%%%%%%%%%%%%%%%%%%%%%%%%%% 

\section{Introduction} 
 
Type IIA orientifolds with intersecting D6-branes and their 
mirror symmetric Type IIB counterparts have 
proven to provide a phenomenologically interesting class of string 
compactifications and have been under intense 
investigation during  the last couple of years
\cite{Uranga:2003pz,Lust:2004ks,Blumenhagen:2005mu,Blumenhagen:2006ci}. 
 
In order to make contact with experiment one needs 
not only the means to determine the  
gauge group and chiral matter content of such a string model, 
but also has to develop tools for the computation of the 
low energy effective action. It is in this latter 
low energy description where  important issues like 
 moduli stabilisation and supersymmetry breaking  
are discussed. 
 
In this paper we would like to clarify certain important aspects 
of this effective action, which to our knowledge have so far 
not been spelled out in the literature. 
The first issue concerns the properties of the  
gauge couplings in $\mathcal{N}=1$ supersymmetric D6-brane vacua.  
The physical one-loop open string threshold corrections to  
the gauge couplings have been computed in \cite{Lust:2003ky, letter} 
for toroidal backgrounds.  
Here we first state a non-renormalisation theorem for the 
holomorphic gauge kinetic function at the one-loop level 
and then explicitly show that it is indeed satisfied 
for the Wilsonian gauge couplings in toroidal Type IIA 
orientifolds.  
 
We revisit perturbative one-loop corrections to the 
Fayet-Iliopoulos (FI) terms for intersecting D6-branes. In \cite{Lawrence:2004sm}  
it was shown that, if the D6-branes are supersymmetric at tree-level, 
in a globally consistent model no such corrections are generated  
at one-loop. We ask the question whether a small non-vanishing FI-term 
on a D6$_b$-brane can induce an FI-term on a D6$_a$-brane which is 
supersymmetric at tree-level. 
We find  the intriguing result that the one-loop 
induced FI-term on brane D6$_a$ can be expressed 
by the gauge threshold corrections. 
     
Moving back to gauge couplings, the extraction of the  
Wilsonian part involves an interesting interplay 
between the non-holomorphic gauge couplings and 
the K\"ahler potentials for all the matter fields involved, 
providing strong constraints on the complete moduli dependence 
of the matter field K\"ahler potentials.  
As we will see, in order to cancel all $\sigma$-model anomalies in the 
effective action, a
one-loop redefinition of the dilaton $S$-field as well as of the complex 
structure moduli $U$ is needed. The hereby induced corrections to the gauge coupling
constants will be referred to as ``universal'' threshold corrections (in analogy
with the heterotic string).
This is in contrast to  perturbative heterotic string compactifications, 
where only the dilaton acquires a universal one-loop field redefinition 
\cite{Derendinger:1991hq,Derendinger:1991kr,LopesCardoso:1991zt,Ibanez:1992hc,Kaplunovsky:1994fg,Kaplunovsky:1995jw}. 
 
Having revisited and discussed the perturbative one-loop corrections,  
in the second part of the paper we undertake some 
first steps towards a better understanding of possible 
D-instanton effects.  
Such effects are  very important for an understanding of 
the vacuum structure (these days called the landscape) 
of string compactifications and it has been pointed out recently 
that they can also generate phenomenologically appealing terms 
like Majorana masses for neutrinos 
\cite{Blumenhagen:2006xt,Ibanez:2006da,Haack:2006cy,Abel:2006yk,Bianchi:2007wy,Cvetic:2007ku,Ibanez:2007rs}. 
Moreover, they are also important for the string theory description 
of gauge instanton effects \cite{Billo:2002hm,Florea:2006si,Akerblom:2006hx,Argurio:2007vq,Bianchi:2007fx} (see \cite{Bianchi:2007ft} for a recent general review on instantons). 
 
String instantons are given by wrapped 
string world-sheets as well as by wrap\-ped Euclidean D-branes and, like in field 
theory, their contributions to the space-time superpotential are  quite 
restricted.  
These contributions can be computed in a semi-classical approach, i.e. 
one involving only the tree-level instanton action and a one-loop 
determinant for the fluctuations around the instanton \cite{Witten:1999eg}. 
For type IIA orientifold models on Calabi--Yau spaces with intersecting 
D6-branes  
(and  their T-dual cousins) the 
contribution of wrapped Euclidean D2-branes, hereafter called E2-branes, to 
the superpotential has been determined in \cite{Blumenhagen:2006xt} 
(see also \cite{Ibanez:2006da}). 
Since both the D6-branes and the E2-instantons are described 
by open string theories,  it was shown that (in the spirit of the D$(-1)$ instantons treated 
e.g. in \cite{Green:1997tv,Gutperle:1997iy,Billo:2002hm}) the entire 
instanton computation boils 
down to the evaluation of disc and one-loop string diagrams with 
boundary (changing) operators inserted.  
Here both the D6-branes and the E2-instantons wrap compact three-cycles 
of the Calabi--Yau manifold.  
 
Intriguingly, the one-loop contributions in the instanton amplitude 
\cite{Blumenhagen:2006xt} 
have been shown to be identical to  
string threshold corrections for the  gauge couplings of the 
corresponding D6-branes \cite{Abel:2006yk,Akerblom:2006hx}. This relates the computation 
of such instanton amplitudes to the discussion in the 
first half of this paper.  
So far it has not been explained explicitly in which sense 
the computed instantonic correlation functions are meant 
to be holomorphic. With the results from the first part 
of this paper, we clarify this point. 
 
Finally, we show that  E2-instantons not only contribute 
to the superpotential but, from the zero mode counting, can  
 also contribute to the holomorphic 
gauge kinetic functions for the $SU(N)$ gauge groups localised 
on the D6-branes. In order for such corrections to arise, 
the E2-instanton must not be rigid but must admit one extra 
pair of fermionic  zero modes arising from a deformation 
of the instanton.   
This is the space-time instanton 
generalisation of a fact known from topological  
string theory, namely that world-sheet instantons 
induce ${\rm tr}(W^2)^{h-1}$ couplings if they have $h$ boundaries. 
We will see that such couplings can also arise 
from space-time E2-instantons.  
Finally, we find that the zero mode counting also allows 
E2-instanton corrections to the FI-terms on the D6-branes.  
Similar to the one-loop corrections, these can 
arise once the supersymmetry on the E2-brane 
is softly broken by for instance turning on the  
$C_3$-form modulus through the world-volume of E2.

\section{A non-renormalisation theorem} 
\label{sectionone} 
 
Let us investigate the structure of perturbative and non-perturbative 
corrections to the holomorphic gauge kinetic functions 
for Type II orientifolds. We  discuss this 
for Type IIA orientifolds, but this is of course 
related via mirror symmetry to the corresponding 
Type IIB orientifolds.  
  
Consider  a Type IIA orientifold with O6-planes 
and intersecting D6-branes preserving ${\cal N}=1$ 
supersymmetry in four dimensions, i.e.  
the D6-branes wrap special Lagrangian (sLag) three-cycles 
$\Pi_a$ of the underlying Calabi--Yau manifold ${\cal X}$, all preserving 
the same supersymmetry. 
On the threefold we introduce in the usual way a symplectic 
basis $(A_I, B^I)$, $I=0,1,\ldots, h_{2,1}$  of homological three-cycles with 
the topological intersection  numbers 
\bea 
             A_I\circ B^J = \delta_{I}^J. 
\eea  
Moreover, we assume that 
the $A_I$ cycles are invariant under the orientifold projection and that the $B^J$ cycles 
are projected out. 
The complexified complex structure moduli on such an orientifold 
are defined as 
\bea 
     U^c_I={1\over (2\pi)\, \ell_s^3}\left[ e^{-\phi_{4}}\,  \int_{A_I} \Re 
       (\widehat\Omega_3)\, -i \int_{A_I} C_3 \right],
\eea 
where $\widehat\Omega_3$ denotes the normalised holomorphic three-form 
on ${\cal X}$ and the four-dimensional dilaton is defined by  
$\phi_4=\phi_{10}-{1\over 2}\ln (V_{\cal X}/\ell_s^6)$. 
Expanding a three cycle $\Pi_a$  into the symplectic basis, 
\bea 
   \Pi_a= M^I_{a}\, A_I + N_{a,I}\, B^I\,, 
\eea 
with $M^I, N_I\in \mbb{Z}$, 
from dimensional reduction of the Dirac-Born-Infeld (DBI) action one can  
deduce the $SU({\cal N}_a)$ gauge kinetic functions at string tree-level  
\bea 
\label{gggff}
           f_a=\sum_{I=0}^{h_{2,1}} M^I_a\, U^c_I. 
\eea 
Since the imaginary parts of the $U^c_I$ are axionic fields, they enjoy 
a Peccei-Quinn shift symmetry $U^c_I\to U^c_I+c_I$ which is preserved 
perturbatively and only broken by E2-brane instantons.  
 
Let $C_i$ denote a basis of anti-invariant 2-cycles, i.e.  
$C_i\in H^{1,1}_{-}$. The complexified K\"ahler moduli are then defined 
as 
\bea 
       T^c_i={1\over \ell_s^2}\left(\int_{C_i} J_2 - i\int_{C_i} B_2\right) \,, 
\eea 
where $B_2$ denotes the NS-NS two-form of the Type IIA superstring. 
Therefore, also the complexified K\"ahler moduli enjoy a 
Peccei-Quinn shift symmetry, broken by world-sheet instantons. 
Note, that the chiral fields $T^c_i$ organise the $\sigma$-model 
perturbation theory and do not contain the dilaton,  
so that the string perturbative theory is entirely defined by powers 
of the $U^c_I$. Moreover, to shorten the notation
we denote by $U^c_I$ and $T^c_i$ the complexified moduli
and by $U_I$ and $T_i$ only the real parts.
 
The superpotential $W$ and the gauge kinetic function $f$ in the 
four-dimensional effective supergravity action are holomorphic 
quantities. In the usual way, employing  holomorphy and 
the Peccei-Quinn symmetries above, one arrives at the following 
two non-renormalisation theorems. 
 
The superpotential can only have the following dependence on $U^c_I$  
and $T^c_i$ 
\bea 
\label{supernon} 
       W=W_{\rm tree} + W^{\rm np}\left(e^{-U^c_I}, e^{-T^c_i} \right),  
\eea 
i.e. beyond tree-level there can only be non-perturbative  
contributions from world-sheet and E2-brane instantons. 
Similarly, the holomorphic gauge kinetic function must look like 
\bea 
\label{gaugenon} 
    f_a=\sum_I M^I_a U^c_I + f_a^{\text{1-loop}}\left(e^{-T^c_i}\right) 
      + f_a^{\rm np}\left(e^{-U^c_I}, e^{-T^c_i} \right),  
\eea 
i.e. in particular its one-loop correction must not depend 
on the complex structure moduli. 
Finally, we consider the Fayet-Iliopoulos terms for the 
$U(1)_a$ gauge fields on the D6-branes. 
At string tree-level and for small deviations from the supersymmetry 
locus, these are given by 
\bea 
    \xi_a= e^{-\phi_4} \,\int_{\Pi_a} \Im (\widehat\Omega_3)=  
       e^{-\phi_4} \, N^I_a \, \int_{B_I} \Im (\widehat\Omega_3) 
\eea 
and therefore only depend on the complex structure moduli. 
At this classical level there are  no $\alpha'$ 
corrections.  
It is an important question about brane stability whether  
these FI-terms receive perturbative or non-perturbative 
corrections in $g_s$. Again, non-renormalisation theorems say that 
in the Wilsonian sense one expects perturbative corrections at most at one-loop.
 
In the following, we will be concerned with the 
terms beyond tree-level appearing in (\ref{supernon}), (\ref{gaugenon}) 
and for the FI-terms. 
First, we discuss the one-loop threshold corrections 
$f^{1-{\rm loop}}\bigl(e^{-T^c_i}\bigr)$, which also make 
their appearance in the space-time instanton generated  
superpotential  $W^{\rm np}\bigl(e^{-U^c_I}, e^{-T^c_i} \bigr)$. 
Second, we will revisit the computation of stringy one-loop 
corrections to the FI-terms. 
Finally, we will discuss $f^{\rm np}\bigl(e^{-U^c_I}, e^{-T^c_i} \bigr)$ 
as well as instanton corrections to the FI-terms. 
 
%%%%%%%%%%%%%%%%%%%%%%%%%%%%%%%%%%%%%%%%%%%%%%%%% 
%%%%%%%%%%%%%%%%%%%%%%%%%%%%%%%%%%%%%%%%%%%%%%%%% 
%%%%%%%%%%%%%%%%%%%%%%%%%%%%%%%%%%%%%%%%%%%%%%%%% 
%%%%%%%%%%%%%%%%%%%%%%%%%%%%%%%%%%%%%%%%%%%%%%%%% 
%%%%%%%%%%%%%%%%%%%%%%%%%%%%%%%%%%%%%%%%%%%%%%%%% 
%%%%%%%%%%%%%%%%%%%%%%%%%%%%%%%%%%%%%%%%%%%%%%%%% 
%%%%%%%%%%%%%%%%%%%%%%%%%%%%%%%%%%%%%%%%%%%%%%%%% 
%%%%%%%%%%%%%%%%%%%%%%%%%%%%%%%%%%%%%%%%%%%%%%%%% 
%%%%%%%%%%%%%%%%%%%%%%%%%%%%%%%%%%%%%%%%%%%%%%%%% 
%%%%%%%%%%%%%%%%%%%%%%%%%%%%%%%%%%%%%%%%%%%%%%%%% 
%%%%%%%%%%%%%%%%%%%%%%%%%%%%%%%%%%%%%%%%%%%%%%%%% 
\section{One-loop thresholds for intersecting D6-branes on $\mbb T^6$} 
The purpose of this section is to recall the one-loop results for the gauge threshold 
corrections in intersecting D6-brane models \cite{Lust:2003ky,Anastasopoulos:2006hn,letter}. 
The gauge coupling constants of the various gauge group factors $G_a$ 
in such a model, 
up to one loop, have the form 
\bea 
\label{thresa} 
 \frac{8\pi^2}{g_a^2(\mu)} = \frac{8\pi^2}{g_{a,\mathrm{string}}^2} + 
                     \frac{b_a}{2}\, \ln\left( \frac{M_s^2}{\mu^2}\right) + \frac{\Delta_a}{2} \,,
 \label{gceffST}  
\eea 
where $b_a$ is the beta function coefficient. The first 
term corresponds to the gauge coupling constant at the string scale, which contains 
the tree-level gauge coupling as well as the ``universal'' contributions
at one-loop (see section 5.2). These contributions are universal in the sense that they originate
from a redefinition of the dilaton and complex structure moduli at one-loop. The redefinition
is brane stack and therefore gauge group independent
. However, as the gauge couplings
differ for the various gauge groups already at tree level, this correction effectively
is gauge group dependent.
The second term gives the usual one-loop 
running of the coupling constants, and the third term denotes the one-loop string threshold 
corrections originating from integrating 
out massive string excitations.  The last two terms can be computed as a 
sum of all annulus and M\"obius diagrams with one boundary on brane $a$ in the presence 
of a background magnetic field in the four-dimensional space-time: 
\bea 
 b_a\, \ln\left( \frac{M_s^2}{\mu^2}\right) + \Delta_a 
 &=& \sum_b T^A({\rm D6}_a,{\rm D6}_b) + \sum_{b'} T^A({\rm D6}_a,{\rm D6}_{b'}) \\ 
 &&  +  \ T^A({\rm D6}_a,{\rm D6}_{a'}) + T^M({\rm D6}_a,{\rm O6}).\nonumber 
\eea 
Here, ${\rm D6}_{c'}$ denotes the orientifold image of brane $c$. In an orbifold one also has to take 
into account the orbifold images of the branes and orientifold planes. 
 
The relevant amplitudes for the $\mbb Z_2 \times \mbb Z_2$ orbifold have been computed 
\cite{Lust:2003ky,letter}. In a sector preserving ${\cal N}=1$ supersymmetry (this means in particular 
$\sum_I \theta_{ab}^I = 0$) the annulus and M\"obius amplitudes are (after subtracting terms which upon 
summing over all diagrams vanish due to the tadpole cancellation condition) \cite{letter} 
\begin{multline} 
T^A({\rm D6}_a,{\rm D6}_b) 
= \frac{I_{ab} N_b}{2}\Biggl[\ln\left( \frac{M_s^2}{\mu^2}\right) 
\sum_{I=1}^3 \sign(\theta_{ab}^I) - \\ 
- \ln \prod_{I=1}^3 
\left( \frac{\Gamma(|\theta_{ab}^I|)}{\Gamma(1-|\theta_{ab}^I|)} \right)^{\sign(\theta_{ab}^I)} 
- \sum_{I=1}^3 \sign(\theta_{ab}^I)\,(\ln 2-\gamma)\Biggr], 
\label{annu4abe3} 
\end{multline} 
\begin{multline} 
T^M({\rm D6}_a,{\rm O6_k})  = 
 \pm I_{a;O6_k} \Biggl[ 
 \ln \left(\frac{M_s^2}{\mu^2}\right) \sum_{I=1}^3\sign(\theta_{a;O6_k}^I) - \\ 
 - \ln \prod_{I=1}^3 \left(\frac{\Gamma(2|\theta_{a;O6_k}^I|)} 
                  {\Gamma(1-2|\theta_{a;O6_k}^I|)}\right)^{\sign(\theta_{a;O6_k}^I)}+  
\sum_{I=1}^3 \sign(\theta_{a;O6_k}^I) (\gamma-3\ln2) \Biggr], 
\label{moe4ae} 
\end{multline} 
where $I_{ab}$ is the intersection number of branes $a$ and $b$, $N_b$ 
is the number of branes on stack $b$ and $\pi \theta_{ab}^I$ is the intersection 
angle of branes $a$ and $b$ on the $I$'th torus. Similarly, $I_{a;O6_k}$ denotes the 
intersection number of brane $a$ and orientifold plane $k$ and $\pi \theta_{a;O6_k}^I$ 
their intersection angle. 
The formula for $T^M$ is only valid for $|\theta_{a;O6_k}^I|<1/2$, the formulas for 
other cases look similar \cite{letter}.

In a sector preserving ${\cal N}=2$ supersymmetry one finds \cite{Lust:2003ky} 
\begin{equation} 
T^A({\rm D6}_a,{\rm D6}_b)= 
 N_b |I_{ab}^J\,  I_{ab}^K| 
 \left[ \ln \left( \frac{M_s^2}{\mu^2}\right) -\ln \vert \eta(i\,T^c_I)\vert^4  
- \ln (T_I\, V_I^a) + \gamma_E - \ln (4\pi) \right], 
\label{annu8ab3} 
\end{equation} 
where $I$ denotes the torus on which the branes lie on top of each other, 
$T_I$ its K\"ahler 
modulus and $T^c_I$ its complexification with $T_I=\Re(T^c_I)$. 
Furthermore, $V_I^a=|n_a^I+iu_Im_a^I|^2/u_I$, with $u_I$ the 
complex structure modulus of the torus and $n_a^I$, $m_a^I$ the wrapping numbers on the $I$'th torus. 
Note that the moduli dependence of the one-loop threshold function in the ${\cal N}=2$ sectors 
is in complete agreement with the non-renormalisation theorems of section two  
(see eq. (\ref{gaugenon})), since 
the holomorphic part of $\Delta_a^{{\cal N}=2}$ is proportional only to $\ln\eta(i\, T^c_I)$. 
 
For   the ${\cal N}=1$ sectors, the one-loop thresholds in a given open string  
D6-brane sector have the following form 
(specialising to the case 
$\theta_{ab}^{1,2}>0$, $\theta_{ab}^3<0$): 
\bea 
\label{thresb} 
\Delta_a=-\frac{b_a}{16\pi^2}\ln\left[ \frac{\Gamma(\theta_{ab}^1) \Gamma(\theta_{ab}^2) \Gamma(1+\theta_{ab}^3)} 
                {\Gamma(1-\theta_{ab}^1) \Gamma(1-\theta_{ab}^2)  
\Gamma(-\theta_{ab}^3)}\right]. 
\eea 
This expression is a non-holomorphic function of the complex structure 
moduli $U_I^c$. Hence, for the ${\cal N}=1$ sectors, 
the holomorphic one-loop gauge kinetic function $f_a^{1-{\rm loop}}\left(e^{-T^c_i}\right)$ 
vanishes. The emergence of the non-holomorphic terms in the one-loop threshold 
corrections will be further discussed in section 5.

%%%%%%%%%%%%%%%%%%%%%%%%%%%%%%%%%%%%%%%%%%%%%%%%% 
%%%%%%%%%%%%%%%%%%%%%%%%%%%%%%%%%%%%%%%%%%%%%%%%% 
%%%%%%%%%%%%%%%%%%%%%%%%%%%%%%%%%%%%%%%%%%%%%%%%% 
%%%%%%%%%%%%%%%%%%%%%%%%%%%%%%%%%%%%%%%%%%%%%%%%% 
%%%%%%%%%%%%%%%%%%%%%%%%%%%%%%%%%%%%%%%%%%%%%%%%% 
%%%%%%%%%%%%%%%%%%%%%%%%%%%%%%%%%%%%%%%%%%%%%%%%% 
%%%%%%%%%%%%%%%%%%%%%%%%%%%%%%%%%%%%%%%%%%%%%%%%% 
%%%%%%%%%%%%%%%%%%%%%%%%%%%%%%%%%%%%%%%%%%%%%%%%% 
%%%%%%%%%%%%%%%%%%%%%%%%%%%%%%%%%%%%%%%%%%%%%%%%% 
%%%%%%%%%%%%%%%%%%%%%%%%%%%%%%%%%%%%%%%%%%%%%%%%% 
%%%%%%%%%%%%%%%%%%%%%%%%%%%%%%%%%%%%%%%%%%%%%%%%% 
%%%%%%%%%%%%%%%%%%%%%%%%%%%%%%%%%%%%%%%%%%%%%%%%% 
\section{Fayet-Iliopoulos terms} 
\label{secfit} 
 
In this section we investigate one-loop corrections to the FI-terms 
for a  $U(1)_a$ gauge field on the D6$_a$-brane induced  
by the presence of other branes D6$_b$. 
Such corrections for Type I string vacua have already been studied 
in \cite{Poppitz:1998dj,Bain:2000fb} and the case of intersecting D6-branes  
has been discussed  in \cite{Lawrence:2004sm}. Here we are following 
essentially the computational technique of \cite{Lawrence:2004sm}. 
The crucial observation is that the vertex operator for the 
auxiliary D-field in the $(0)$-ghost picture is simply given 
by the internal world-sheet $U(1)$ current, i.e.  $V^{(0)}_{D_a}=J_{U(1)}$. 
Therefore, the one-point function of $V^{(0)}_{D_a}$ on 
the annulus with boundaries $a$ and $b$ can  be written as 
\bea 
   \langle V_{D_a} \rangle= -{i\over 2\pi}\, \partial_\nu \int_0^ \infty  dt\,\,  
     Z_{ab}(\nu, it)|_{\nu=0}, 
\eea 
where $Z_{ab}(\nu, it)$ denotes the annulus partition function, 
with insertion of $\exp(2 \pi i J_0)$, in the open string sector $(ab)$, where $J_0$ is the zero mode of the 
$U(1)$ current.  
In the case of intersecting D6-branes on a torus preserving ${\cal N}=1$ 
supersymmetry and after application of the Riemann theta-identities, 
this partition function is given by 
\bea 
   Z_{ab}(\nu, it)= I_{ab}\, N_b \, { (-i)^3\over \pi^4 t^2 }\, {\vartheta_1({3\nu\over 2}, it) 
     \prod_I \vartheta_1(-{\nu\over 2} + i{\theta_I\over 2}t,it )\over 
     \eta^3 (it)\, \prod_I \vartheta_1( i{\theta_I\over 2}t, it ) }. 
\eea 
Using that $\vartheta_1(0, it)=0$ and $\vartheta'_1(\nu,it)|_{\nu=0}=-2\pi\eta^3$, 
one obtains the divergence 
\bea 
   \langle V_{D_a} \rangle\simeq I_{ab}\, N_b \int_0^\infty  {dt\over t^2} , 
\eea 
which is cancelled by tadpole cancellation in a global model. 
Therefore, once the D6-branes are supersymmetric at tree-level, no 
FI-term is generated at one-loop level and the system is not destabilised. 
This result is consistent with the computations in \cite{Poppitz:1998dj,Bain:2000fb}. 
 
However, this is not the end of the story of the one-loop corrections to   
FI-terms in intersecting D6-brane models.  
One can also envision that a tree-level FI-term on a brane D6$_b$ induces 
via a one-loop diagram an FI-term on a  brane D6$_a$. 
To proceed we assume that $\theta^I_b\to \theta^I_b+2\epsilon^I$ with  
$\sum \epsilon^I=\epsilon$  
and compute 
\bea 
   \langle V_{D_a} \rangle_{\epsilon^I}= -{i\over 2\pi}\, \partial_\nu\partial_{\epsilon^I} \int_0^ \infty  dt\,\,  
     Z_{ab}(\epsilon^I,\nu, it)|_{\nu=0, \epsilon^I=0} 
\eea 
with 
\bea 
   Z_{ab}(\epsilon^I, \nu, it)= I_{ab}\, N_b \, { (-i)^3\over \pi^4 t^2 }\,  
   {\vartheta_1({3\nu\over 2}+ i{\epsilon\over 2}t, it) 
     \prod_I \vartheta_1 \bigl( -{\nu\over 2} + i{(\epsilon^I-\epsilon^J-\epsilon^K)\over 2}t  
     + i{\theta_I\over 2}t,it \bigr)\over 
     \eta^3 (it)\, \prod_I \vartheta_1( i{\epsilon^I}t + i{\theta_I\over 2}t, it ) }.\nonumber 
\eea 
The derivative with respect to the supersymmetry breaking parameters $\epsilon^I$ brings 
down one factor of $t$ and it turns out that the result is the same for all $\partial_{\epsilon^I}$: 
\bea 
    \langle V_{D_a} \rangle_{\epsilon}\simeq  i\, I_{ab}\, N_b \, \int_0^\infty  {dt\over t} 
     \sum_{I=1}^3\,\, 
         \frac{\vartheta_1'}{\vartheta_1} \left( {\textstyle \frac{i \theta_{ab}^I t}{2},\frac{it}{2}}\right). 
\eea 
This is the same expression as the one-loop threshold corrections $T^A(D6_a,D6_b)$, so that 
at  linear order in $\epsilon$ we obtain for the FI-terms 
\bea 
     (\xi_a)_\epsilon  = (\epsilon_b-\epsilon_a)\,  T^A(D6_a,D6_b). 
\eea  
Completely analogously, one can show that this formula remains 
true also for ${\cal N}=2$ open string sectors.  
Therefore, we would like to propose that such a relation between gauge threshold 
and one-loop corrections to FI-terms is valid for general intersecting 
D6-brane models. Moreover, the Wilsonian part of the thresholds $T^A(D6_a,D6_b)$, which  
we compute in the next section, should also be the Wilsonian part of the 
correction to the FI-terms. 
%%%%%%%%%%%%%%%%%%%%%%%%%%%%%%%%%%%%%%%%%%%%%%%%%%%% 
%%%%%%%%%%%%%%%%%%%%%%%%%%%%%%%%%%%%%%%%%%%%%%%%%%%% 
%%%%%%%%%%%%%%%%%%%%%%%%%%%%%%%%%%%%%%%%%%%%%%%%%%%% 
%%%%%%%%%%%%%%%%%%%%%%%%%%%%%%%%%%%%%%%%%%%%%%%%%%%% 
%%%%%%%%%%%%%%%%%%%%%%%%%%%%%%%%%%%%%%%%%%%%%%%%%%%% 
%%%%%%%%%%%%%%%%%%%%%%%%%%%%%%%%%%%%%%%%%%%%%%%%%%%% 
%%%%%%%%%%%%%%%%%%%%%%%%%%%%%%%%%%%%%%%%%%%%%%%%%%%% 
%%%%%%%%%%%%%%%%%%%%%%%%%%%%%%%%%%%%%%%%%%%%%%%%%%%% 
 
\section{Wilsonian gauge kinetic function and $\sigma$-model anomalies}

In a  supersymmetric gauge theory one can compute the running 
gauge couplings $g_a(\mu^2)$ in terms of the gauge kinetic functions $f_a$, the 
K\"ahler potential ${\cal K}$ and the K\"ahler metrics of the charged matter 
fields $K^{ab}(\mu^2)$ 
\cite{Shifman:1986zi,Derendinger:1991hq,Derendinger:1991kr,LopesCardoso:1991zt,Ibanez:1992hc,Kaplunovsky:1994fg,Kaplunovsky:1995jw}: 
\bea 
\label{gceffFT} 
 \frac{8\pi^2}{g_a^2(\mu^2)} &=& 8\pi^2\,\Re(f_a)+ \frac{b_a}{2}\,  
 \ln \left( \frac{\Lambda^2}{\mu^2}\right) 
   + \frac{c_a}{2} \, {\cal  K} \nonumber \\ && 
   + T(G_a) \ln g_a^{-2}(\mu^2)  
   - \sum_r T_a(r) \ln \det K_r (\mu^2), 
\eea 
with 
\bea 
\label{betafctcoeff} 
b_a =  \sum_r n_r T_a(r) - 3\,  T(G_a), \quad\quad  
c_a= \sum_r n_r T_a(r) - T(G_a) 
\eea 
and $T_a(r) = {\rm Tr} (T^2_{(a)})$ ($T_{(a)}$ being the 
generators of the gauge group $G_a$). 
In addition, $T(G_a) = T_a({\rm adj}\, G_a)$ and $n_r$ is the 
number of multiplets in the representation $r$ of the gauge group and the  
sums run over these representations.  
In this context, the natural cutoff scale for a field theory is the Planck scale, 
i.e. $\Lambda^2=M_{\mathrm{Pl}}^2$.  
 
The left hand side of eq. \eqref{gceffFT} is given by eq. \eqref{thresa}, which contains the gauge coupling 
at the string scale, $1/g_{a,\mathrm{string}}^2$, as well as 
the one-loop string threshold 
corrections $\Delta_a$. 
In general, $\Delta_a$ is the sum of a non-holomorphic term plus the real-part of a holomorphic 
threshold correction: 
\bea  
\Delta_a=\Delta_{a}^{\rm n.h.}+\Re(\Delta_{a}^{\mathrm{hol.}})\, . 
\eea

On the right hand side of eq. (\ref{gceffFT}), $f_a$ denotes the Wilsonian, i.e. holomorphic, gauge kinetic function, 
which is given in terms of a holomorphic tree-level function plus the holomorphic part of 
the one-loop threshold corrections (cf. eq. (\ref{gaugenon})): 
\bea 
\label{gaugenona} 
    f_a=f_a^{\rm tree}+ f_a^{1-{\rm loop}}\left(e^{-T^c_i}\right)= 
\sum_I M^I_a U^c_I +\Delta_{a}^{\mathrm{hol.}}\, . 
\eea 
In addition, on the right hand side of eq. (\ref{gceffFT}) 
the non-holomorphic terms proportional to the 
K\"ahler metric of the moduli ${\cal K}$ and the matter field K\"ahler metrics $K_r$ are 
due to the one-loop contributions of massless fields. These fields generate non-local terms 
in the one-loop effective action, which correspond to one-loop non-invariances 
under $\sigma$-model transformations, the so-called $\sigma$-model  anomalies (K\"ah\-ler  
and reparametrisation anomalies).

Matching up all terms in eq. \eqref{gceffFT} essentially means that 
the $\sigma$-model anomalies can be cancelled in a two-fold way. 
First, by local contributions to the gauge coupling constant 
via the one-loop threshold contributions $\Delta_a$. 
These terms originate from massive string states. 
The second way to cancel the $\sigma$-model anomalies is due to a field 
dependent (however gauge group independent)
 one-loop contribution to the K\"ahler potential of the chiral 
moduli fields. It implies that some of the moduli fields transform non-trivially 
under the K\"ahler transformations and also under reparametrisations 
in moduli space. The universal one-loop modification of the K\"ahler potential  is nothing else 
than a generalised Green-Schwarz mechanism cancelling the $\sigma$-model 
anomalies. This is analogous to the Green-Schwarz mechanism which cancels anomalies 
of physical $U(1)$ gauge fields, whereas the $\sigma$-model anomalies correspond to 
unphysical, composite gauge connections. 
Effectively it means that the Green-Schwarz mechanism with respect to the $\sigma$-model 
anomalies can be described by a non-holomorphic, one-loop  field redefinition of 
the associated tree-level moduli fields.  

As we will see, in type IIA orientifold models these field redefinitions act on the real parts of the 
dilaton field $S$ as well as the complex structure moduli $U_J$: 
\bea 
\label{gs} 
 S &\rightarrow& S +\delta^{GS}(U,T)  \nonumber \\ 
 U_J &\rightarrow& U_J + \delta_J^{GS}(U,T). 
\eea 
These redefined fields are those that determine the gauge coupling constants  
$1/g_{a,\mathrm{string}}^2$ at the string scale. 
Recall that,
as the tree-level gauge couplings (\ref{gggff}) are already gauge group dependent,
so are these one-loop corrections, but the only dependence
arises due to the universal one-loop redefinition of the moduli
fields. It is in this sense that 
we still call these one-loop corrections
to the gauge couplings universal.   
Note that in 
heterotic string compactifications, the $\sigma$-model 
Green-Schwarz mechanism only acts on the heterotic dilaton field.

\subsection{Holomorphic gauge couplings for toroidal models} 
 
In summary,  
equation (\ref{gceffFT}) is to be understood recursively, which 
means that one can insert the tree-level 
results into  the last three terms of eq. (\ref{gceffFT}). In addition, one also has to include  
the universal field redefinition eq. (\ref{gs}) in $1/g_{a,\mathrm{string}}^2$ in the 
left hand side of eq. (\ref{gceffFT}), in order to get a complete matching of all terms in  
eq. (\ref{gceffFT}), as 
we will demonstrate for the aforementioned toroidal orbifold 
in the following. For ${\cal N}=2$ sectors the one-loop threshold corrections 
to the gauge coupling constant indeed contain a holomorphic, Wilsonian term $f_a^{(1)}$, 
whereas for ${\cal N}=1$ sectors only the non-holomorphic piece $\Delta_a^{\rm n.h.}$ is present. 
 
Specifically, 
the holomorphic gauge kinetic function can now be determined by comparing the string 
theoretical formula (\ref{gceffST}) for the effective gauge 
coupling with the field theoretical one (\ref{gceffFT}). The first thing to notice are the 
different cutoff scales appearing in the two formulas. One needs to convert one into the other 
using 
\bea 
\frac{M_s^2}{M_\mathrm{Pl}^2}\propto \exp (2\phi_4) \propto  (S\, U_1\, U_2\, U_3)^{-\frac{1}{2}}.\label{msmprel} 
\eea 
Here, $\phi_4$ is the four dimensional dilaton and the complex 
structure moduli in the supergravity basis can be expressed 
in terms of $\phi_4$ and the complex structure moduli 
$u_I=R_{I,2}/R_{I,1}$ as 
\bea 
   S={1\over 2\pi}\, e^{-\phi_4}\, {1\over \sqrt{u_1\, u_2\, u_3}}, \quad 
   U_I={1\over 2\pi}\, e^{-\phi_4}\, \sqrt {u_J\, u_K \over u_I},\ {\rm with}\ I\ne J\ne K \ne I\, . 
\eea 
These fields are  the real parts of complex scalars of four dimensional chiral 
multiplets $S^c$ and $U_I^c$.

As $\mathcal{N}=4$ super Yang--Mills theory is finite, one expects the sum of the terms 
in (\ref{gceffST}) proportional to  $T(G_a)$ to cancel. This is because the only chiral multiplets  
transforming in the adjoint representation of the gauge group are the open string 
moduli which (on the background considered) assemble themselves into three chiral multiplets, thus forming an 
$\mathcal{N}=4$ sector together with the gauge fields. To show that this cancellation does happen, 
one notices the following. Firstly, $n_{\mathrm{adjoint}}=3$, as explained, such that there is no term in $b_a$ 
proportional to $T(G_a)$. Secondly, 
\bea 
{\cal K} &=& -\ln (S^c+\ov S^c) -\sum_{I=1}^3 \ln (U^c_I+\ov U^c_I) - \sum_{I=1}^3 \ln
(T^c_I+\ov T^c_I)  \label{kaehlerpot} \\\label{treekp} 
g_{a,{\rm tree}}^{-2} &=& S\, \prod_{I=1}^3 n_a^I - \sum_{I=1}^3 U_I\,   
n_a^I m_a^J m_a^K 
    \qquad I \neq J \neq K \neq I, 
\eea 
where $T_I$ are the K\"ahler moduli of the torus and $n_a^I$, $m_a^I$  
are the wrapping numbers of the brane.  
Finally, one needs the matter metric for 
the open string moduli, which can be obtained from the T-dual expression in 
models with D9- and D5-branes \cite{Blumenhagen:2006ci}. Performing the 
T-duality, which essentially amounts to exchanging K\"ahler and complex structure moduli 
and converting gauge flux into non-trivial intersection angles for the D6-branes, one arrives 
at ($I=1,\hdots,3$)\footnote{An overall factor involving the wrapping numbers was introduced in this 
expression in order to achieve full cancellation. This can be done, as the expressions used are 
derived only up to overall constants \cite{Lust:2004cx}.}: 
\bea 
 K_{ij}^I = \frac{\delta_{ij}}{T_I U_I} \, 
         \Biggl\vert  
         \frac{(n_a^J + i u_J\, m_a^J)(n_a^K + i u_K\, m_a^K)}{(n_a^I + i u_I\, m_a^I)} 
         \Biggr\vert \qquad I\neq J \neq K  \neq I. 
\eea 
 
Let us now turn to the fields in the fundamental representation of the gauge group $G_a$, 
in particular to the fields arising from the intersection with one other stack of branes, 
denoted by $b$. 
For an ${\cal N}=1$ open string sector the metric for these fields  
can be written as \cite{Lust:2004cx,Blumenhagen:2006ci} (see also \cite{Kors:2003wf}) 
\bea 
\label{kaehlermetricf} 
 K^{ab}_{ij} = \delta_{ij}\, S^{-\alpha}  
     \prod_{I=1}^3 U_I^{-(\beta + \xi\, \theta_{ab}^I)}\  T_I^{-(\gamma + \zeta\, \theta_{ab}^I)} 
     \sqrt{\frac{ 
                \Gamma(\theta_{ab}^1) \Gamma(\theta_{ab}^2) 
                \Gamma(1+\theta_{ab}^3) 
                }{ 
                \Gamma(1-\theta_{ab}^1) \Gamma(1-\theta_{ab}^2) 
                \Gamma(-\theta_{ab}^3) 
                }},  
\eea 
where $\alpha$, $\beta$, $\gamma$, $\xi$ and $\zeta$ are undetermined constants. 
As $\theta_{ab}^{1,2} > 0$ and $\theta_{ab}^{3} < 0$, which is assumed in  
\eqref{kaehlermetricf}, the intersection number $I_{ab}$ 
is positive, implying that 
\bea 
n_f = I_{ab} N_b.\label{numberf} 
\eea 
Using $T_a(f)=\frac{1}{2}$ and relations 
(\ref{betafctcoeff}, \ref{msmprel}, \ref{kaehlerpot}, \ref{kaehlermetricf}, \ref{numberf}) 
one finds a contribution to the right hand side  of (\ref{gceffFT})  
proportional to 
\bea 
&& \frac{I_{ab} N_b}{2} \Biggl( \ln\left( \frac{M_s^2}{\mu^2}\right) +  
(2 \gamma -1) \ln( T_1 T_2 T_3 ) 
+ (2 \beta - {\textstyle \frac{1}{2}}) \ln (U_1 U_2 U_3) +  
(2 \alpha - {\textstyle \frac{1}{2}}) \ln S \nonumber \\ 
&&+ \zeta \sum_{I=1}^3 \theta_{ab}^I \ln T_I +  
 \xi \sum_{I=1}^3 \theta_{ab}^I \ln U_I 
- \ln\left[  \frac{ 
                \Gamma(\theta_{ab}^1) \Gamma(\theta_{ab}^2) 
                \Gamma(1+\theta_{ab}^3) 
                }{ 
                \Gamma(1-\theta_{ab}^1) \Gamma(1-\theta_{ab}^2) 
                \Gamma(-\theta_{ab}^3)}\right] 
                 \Biggr). 
\eea 
Using (\ref{thresb}) one finds that the first and the last term exactly reproduce the 
contribution of the last two terms in (\ref{gceffST}). 
The terms proportional to $\zeta$ and $\xi$ will later be shown to 
constitute the aforementioned universal gauge coupling correction.
The remaining three terms can neither be attributed to such a correction nor can they be written 
as the real part of a holomorphic function. Thus they cannot be the one-loop correction to 
the gauge kinetic function and  therefore must vanish. This fixes some of  
the coefficients in the ansatz (\ref{kaehlermetricf}): 
\bea 
\alpha = \beta = \frac{1}{4} \,, \quad \gamma = \frac{1}{2}. 
\label{fixedcoeffs} 
\eea 
The same matching of terms appears between the M\"obius diagram plus the annulus with boundaries 
on brane $a$ and its orientifold image and the K\"ahler metrics for fields in the symmetric and antisymmetric 
representation. Here, one has to replace $\theta_{ab}^I$ and $I_{ab} N_b$ by $\theta_{aa'}^I=2\theta_{a}^I$ 
and $I_{aa'} N_a$ in \eqref{kaehlermetricf} and \eqref{annu4abe3}. Apart from these replacements, 
the K\"ahler metric for matter in these representations is also given by \eqref{kaehlermetricf} 
with the constants $\alpha$, $\beta$, $\gamma$ given in \eqref{fixedcoeffs}.

The corrections to the gauge couplings coming from ${\cal N}=2$ open string  
sectors  were seen in the previous section to take on quite a different form.  
They contain a term, 
\bea 
  -\ln \left\vert \eta(i\, T^c_I)\right\vert^4 = - 4\, {\rm Re} 
 \left[ \ln \eta(i\, T^c_I)\right],  
\eea 
which can be written as the real part of a holomorphic function. This leads one to 
conclude that the gauge kinetic function receives one-loop corrections from these 
sectors. Inserting the correct prefactor, which from the first term in (\ref{annu8ab3}) 
and the corresponding one in (\ref{gceffFT}) can be seen to be proportional to 
the beta function coefficient, gives 
\bea 
 f_a^{(1)} = - \frac{N_b\,  |I_{ab}^J\,  I_{ab}^K|}{4\pi^2} \,\,  \ln \eta(i\,
 T^c_I) 
 \qquad I \neq J \neq K \neq I, 
\eea 
where again $I$ denotes the torus in which the branes lie on top of each other and 
$I_{ab}^{J,K}$ are the intersection numbers on the other tori.

The term $- \ln ( T_I\, V_I^a)$ in (\ref{annu8ab3}) is not the real part of a 
holomorphic function. Proceeding as before, one finds that the K\"ahler 
metric for the hypermultiplet (or two chiral multiplets) living at 
an intersection of branes $a$ and $b$ preserving eight supercharges 
must be 
\bea 
 K_{ij}^I = \frac{|n_a^I + i u_I\, m_a^I|}{(U^J\, U^K\, T^J\, T^K)^{\frac{1}{2}}}  
 \qquad I \neq J \neq K \neq I. 
\eea 
Apart from the factor in the numerator, this is in agreement with the form found 
by direct calculations \cite{Lust:2004cx,Blumenhagen:2006ci}. The appearance 
of the numerator  is however 
plausible as it also appears in the open string moduli metric and the hypermultiplets 
under discussion should feel the $I$'th torus in the same way. 
 
\subsection{Universal threshold corrections} 
 
In the following, the aforementioned ``universal'' gauge coupling corrections will be 
discussed. They also appear in the heterotic \cite{ Kaplunovsky:1995jw} and type I 
\cite{Antoniadis:1999ge,Berg:2004ek} string and are related to a redefinition of the dilaton 
at one-loop \cite{Derendinger:1991kr,Derendinger:1991hq}. This stems from the fact 
that the dilaton really lives in a linear multiplet rather than a chiral one. 
 
Our general ansatz for the K\"ahler metrics for the chiral matter in an ${\cal N}=1$  sector  
contains a factor 
\bea 
 \prod_{J=1}^3 U_J^{-\xi\, \theta_{ab}^J}\,\, T_J^{-\zeta\, \theta_{ab}^J}, 
 \label{matmetangle} 
\eea 
which according to (\ref{gceffFT}) appears in the one-loop correction to the 
gauge coupling constant. Neither is this term reproduced in the string one-loop 
calculation of the coupling nor can it be written as a correction to the 
holomorphic gauge kinetic function. Therefore, as is familiar from gauge threshold 
computations, there remains the possibility that  
it can be absorbed into a  
one-loop correction to  the $S$ and $U_I$ chiral superfields.  
In the following, we require that such a gauge group factor independent 
universal correction is possible and see how this 
fixes  the parameters in (\ref{matmetangle}). 
 
The first observation is, that in order to get something gauge group 
independent, the factor (\ref{matmetangle}) 
actually must have the following form 
\bea 
 \prod_{J=1}^3 U_J^{-\xi'\, \sign(I_{ab}) \theta_{ab}^J}\,\,  
 T_J^{-\zeta'\, \sign(I_{ab}) \theta_{ab}^J} \,, 
 \label{matmetanglep} 
\eea 
with $\xi'$ and $\zeta'$ independent of the brane. For the metrics of fields 
transforming in the symmetric or antisymmetric representation of the gauge group, 
one has to replace $\phi_{ab}=\phi_a-\phi_b$ by $\phi_{aa'}=2\phi_{a}$ and $\sign(I_{ab})$ by 
$\sign(I_{aa'}-I_{a;O6})$ or $\sign(I_{aa'}+I_{a;O6})$, respectively. Then one computes 
($K'$ denotes the factor 
(\ref{matmetangle}, \ref{matmetanglep}) appearing in the full K\"ahler metric $K$): 
\bea 
\label{gcuniversal} 
 && \sum_{r} T_a(r) \ln \det {{K'}^r} =  
 {\textstyle \frac{|I_{ab}|N_b}{2}} \ln\left[ \prod_{J=1}^3  
        U_J^{-\xi' \sign(I_{ab})\, \theta_{ab}^J}\,\, T_J^{-\zeta' 
 \sign(I_{ab}) \,\theta_{ab}^J}\right] \\  
 && \phantom{aaaaaaaaaaaaaaaai} +{\textstyle \frac{|I_{ab'}|N_b}{2}} \ln\left[ \prod_{J=1}^3  
        U_J^{-\xi' \sign(I_{ab'})\, \theta_{ab'}^J}\,\, T_J^{-\zeta' \sign(I_{ab'})\, 
                  \theta_{ab'}^J}\right] \nonumber \\  
 && \phantom{aaaaaaaa} +{\textstyle \frac{N_a+2}{2} \frac{|I_{aa'}-I_{a;O6}|}{2}} \ln\left[ \prod_{J=1}^3  
        U_J^{-2 \xi' \sign(I_{aa'}-I_{a;O6})\, \theta_{a}^J}\,\, T_J^{-2 \zeta' \sign(I_{aa'}-I_{a;O6}) 
                  \, \theta_{a}^J}\right] 
        \nonumber \\ 
 && \phantom{aaaaaaaa} +{\textstyle \frac{N_a-2}{2} \frac{|I_{aa'}+I_{a;O6}|}{2}} \ln\left[ \prod_{J=1}^3  
        U_J^{-2 \xi' \sign(I_{aa'}+I_{a;O6})\, \theta_{a}^J}\,\,  T_J^{-2 \zeta' \sign(I_{aa'}+I_{a;O6}) 
              \, \theta_{a}^J}\right]\, . 
        \nonumber  
\eea 
After a few steps, using $|I_{ab}|\, \sign(I_{ab})=I_{ab}$  and  
the tadpole cancellation condition,  
this can be brought to the simple form 
\bea 
\label{gcuniversal2} 
\sum_{r} T_a(r) \ln \det {{K'}^r}  &=& - n_a^1 n_a^2 n_a^3\left[ \sum_b N_b\, m_b^1 m_b^2 m_b^3\,  
               \sum_{I=1}^3 \theta_b^I\, (\xi' \ln U_I + \zeta' \ln T_I)\right] \nonumber \\ 
 &-& \sum_{J=1}^3 n_a^J m_a^K m_a^L \left[\sum_b N_b m_b^J n_b^K n_b^L \, 
               \sum_{I=1}^3 \theta_b^I\, (\xi' \ln U_I + \zeta' \ln T_I)\right] \nonumber \\ 
 && \hspace{150pt} J\neq K \neq L\neq J \, . 
\eea 
Therefore, these corrections have precisely the form required for them to be 
identified with the one-loop correction between the 
linear superfields appearing in string theory and the chiral 
superfields used in the supergravity description 
\bea 
\label{sigmatrans} 
 S^{L} &=& S - \frac{1}{8\pi^2} \sum_b N_b\, m_b^1 m_b^2 m_b^3\,  
               \sum_{I=1}^3 \phi_b^I\, (\xi' \ln U_I + \zeta' \ln T_I) \nonumber \\ 
 U^{L}_J &=& U_J + \frac{1}{8\pi^2} \sum_b N_b\, m_b^J n_b^K n_b^L\,  
               \sum_{I=1}^3 \phi_b^I\, (\xi' \ln U_I + \zeta' \ln T_I)  \nonumber \\ 
 && \hspace{200pt} J\neq K \neq L\neq J. 
\eea 
In contrast to eq. (\ref{treekp}), where the tree-level gauge couplings are determined, 
the one-loop gauge couplings at the string scale have to include the redefined 
fields $S^L$ and $U^L$: 
\bea 
g_{a,{\rm string}}^{-2} = S^L\, \prod_{I=1}^3 n_a^I - \sum_{I=1}^3 U_I^L\,   
n_a^I m_a^J m_a^K. 
\eea 
 
In contrast to all (to us)  known cases studied in the
literature, for $\xi'\ne 0$ the fields which are 
corrected, i.e. the moduli $S$ and $U_I$,  also appear in the one-loop redefinition.  
Let us propose an argument, why such corrections might be expected
to be absent: 
Due to the anomalous $U(1)$ gauge symmetries, the 
chiral superfields $S$ and $U_I$ participate in the 
Green-Schwarz mechanism and therefore transform 
non-trivially under $U(1)$ gauge transformations. 
This implies that, in order to be gauge invariant, the one-loop corrections in (\ref{sigmatrans}) 
proportional to $\ln U_I$ 
must be extended in the usual way by $U_I\to U_I +\delta^a_{GS} V_a$. 
Computing the resulting FI-terms via the supergravity formula 
${\xi_a/2g_a^2}=\partial K /\partial V_a |_{V_a=0}$ gives, 
besides the tree-level result depending on $S^L, U^L_I$, a one-loop contribution proportional to 
$\xi'\sum_b \xi_b^{(0)}\phi^J_b /U_J$.  This has an  extra  
dependence on the complex structure moduli $U_I$. However, for intersecting 
D6-branes, we have seen that the Wilsonian (supergravity) FI-Terms 
are proportional to the Wilsonian gauge threshold corrections, which depend only 
on the K\"ahler moduli via instanton corrections (for ${\cal N}=1$ sectors they are  
even vanishing in the setup at hand). 
This seems to suggest that there should better  
be no $U_I^{-\theta_I}$ dependence in the matter field 
K\"ahler metrics, i.e. $\xi'=0$. 

Moreover, in analogy to the heterotic string we expect that
for the range $-1\le \theta_J\le 1$ the exponent of $T_J$ 
runs over the range $[-1,0]$. This condition would fix
$\zeta'=\pm 1/2$.

Let us summarise the conclusions we have  drawn from requiring 
holomorphy of the Wilsonian gauge kinetic function. 
First, we provided arguments that the K\"ahler metric for 
${\cal N}=1$ chiral matter fields  in 
intersecting D6-brane models is of the  following form 
\bea 
\label{kaehlermetrictot} 
 \hspace{-1.0cm}  K^{ab}_{ij} = \delta_{ij}\, S^{-\frac{1}{4}}\,  
     \prod_{J=1}^3 U_J^{-\frac{1}{4}}  
              \,\,     T_J^{-\left(\frac{1}{2} \pm \frac{1}{2}  \sign(I_{ab})\, 
                \theta_{ab}^J\right)}  
     \sqrt{\frac{ 
                \Gamma(\theta_{ab}^1) \Gamma(\theta_{ab}^2) 
                \Gamma(1+\theta_{ab}^3) 
                }{ 
                \Gamma(1-\theta_{ab}^1) \Gamma(1-\theta_{ab}^2) 
                \Gamma(-\theta_{ab}^3) 
                }},  
\eea 
where supersymmetry of course requires $\sum_{I=1}^3 \theta_{ab}^I=0$. 
Second, the holomorphic gauge kinetic function (on the background considered) only 
receives corrections from ${\cal N}=2$ open string sectors 
and the  
one-loop correction takes on the following form 
\bea 
 f_a^{(1)} = - \sum_b \frac{N_b\, |I_{ab}^J I_{ab}^K|}{4\pi^2} \,\, \ln
 \eta(i\, T^c_I) 
 \qquad I \neq J \neq K \neq I, 
\eea 
where the sum only runs over branes $b$ which lie on top of brane $a$ in  
exactly one torus, denoted by $I$. 
Therefore, the results for the gauge threshold corrections 
and the matter field K\"ahler metrics are consistent 
both with the non-renormalisation theorem from section \ref{sectionone} 
and the Kaplunovsky-Louis formula (\ref{gceffFT}). 
Clearly, it would be interesting, along the lines of 
\cite{Lust:2004cx} to carry out a string amplitude computation
to fix the free coefficient in the ansatz (\ref{kaehlermetricf}) and see whether
our indirect arguments are correct.
%%%%%%%%%%%%%%%%%%%%%%%%%%%%%%%%%%%%% 
%%%%%%%%%%%%%%%%%%%%%%%%%%%%%%%%%%%%% 
%%%%%%%%%%%%%%%%%%%%%%%%%%%%%%%%%%%%% 
%%%%%%%%%%%%%%%%%%%%%%%%%%%%%%%%%%%%% 
%%%%%%%%%%%%%%%%%%%%%%%%%%%%%%%%%%%%% 
%%%%%%%%%%%%%%%%%%%%%%%%%%%%%%%%%%%%% 
%%%%%%%%%%%%%%%%%%%%%%%%%%%%%%%%%%%%% 
%%%%%%%%%%%%%%%%%%%%%%%%%%%%%%%%%%%%% 
%%%%%%%%%%%%%%%%%%%%%%%%%%%%%%%%%%%%% 

\section{Holomorphic  E2-instanton amplitudes} 
 
\label{seca}

Space-time instantons are given also by D-branes, which 
in this case are Euclidean D2-branes (so-called 
E2-branes) wrapping three-cycles $\Xi$ in the Calabi--Yau, so that they 
are point-like in four-dimensional 
Minkowski space. Such instantons can contribute 
to the holomorphic superpotential and gauge kinetic 
functions only  if they preserve half of the ${\cal N}=1$ supersymmetry. 
This means that the instanton measure must contain a factor  
$d^4 x\, d^2\theta$. 
Let us first clarify an important aspect of this half-BPS 
condition.  
In the second  part of this  section we then revisit the computation  
of contributions 
of such instantons to the superpotential and also clarify some 
issues concerned with the appearing one-loop determinants. 
In the third and fourth part, we investigate under which conditions 
such string instantons can also contribute to the gauge kinetic functions and FI-terms. 
  
\subsection{Half-BPS instantons} 
 
As has been explained  in \cite{Argurio:2007vq,Bianchi:2007wy,Ibanez:2007rs},  
just wrapping an E2-instanton around a rigid sLag three-cycle 
in the Calabi-Yau gives four bosonic and four 
fermionic zero modes. The vertex operators for the latter 
are 
\bea 
 V^{(-1/2)}_{\theta}(z)= \theta_\alpha\ e^{-{\varphi(z)\over 2}}(z)\, S^\alpha(z)\, \, 
      \Sigma_{h={3\over 8},q={3\over 2}}(z)\ 
\eea 
and 
\bea 
 V^{(-1/2)}_{\ov\theta}(z)=\ov\theta_{\dot\alpha}\ e^{-{\varphi(z)\over 2}}\, {S}^{\dot\alpha}(z)\,\,  
      \Sigma_{h={3\over 8},q=-{3\over 2}}(z)\,. 
\eea 
Therefore, if the instanton is not invariant under 
the orientifold projection, one still has four instead 
of the desired two fermionic zero modes. 
Thus, only by placing the E2-brane in a position 
invariant under $\Omega\ov\sigma$ does one have a chance 
to get rid of the two additional zero modes $\ov\theta$. 
For so called $O(n)$ instantons one can see that 
the zero modes $x_\mu,\theta$ are symmetrised and the  
mode $\ov\theta$ gets anti-symmetrised.  
For the opposite projection, i.e. for $USp(2n)$ instantons, 
the zero modes $x_\mu,\theta$ are anti-symmetrised and the  
mode $\ov\theta$ gets symmetrised.  
Therefore, one can only get the simple $d^4 x\, d^2\theta$ instanton 
measure for a single $O(1)$ instanton.

\subsection{Superpotential contributions}

In order to contribute to the superpotential,  
we also require that there do not arise 
any further zero modes from E2-E2 open strings, 
so that the three-cycle $\Xi$ should be rigid, i.e. $b_1(\Xi)=0$. 
Therefore, considering an E2-instanton in an intersecting brane configuration,  
additional zero modes can only arise 
from the intersection of the instanton $\Xi$ with D6-branes $\Pi_a$. 
There are  $N_a\, [\Xi\cap \Pi_a]^+$ chiral fermionic 
zero modes $\lambda_{a,I}$ and $N_a\, [\Xi\cap \Pi_a]^-$ 
anti-chiral ones, $\ov{\lambda}_{a,J}$.\footnote{Here we introduced  
the physical intersection number between two branes $\Pi_a\cap\Pi_b$, which  
is the sum of positive $[\Pi_a\cap \Pi_b]^+$ and negative  
$[\Pi_a\cap \Pi_b]^-$ intersections.}

For its presentation it is useful to  introduce the short-hand notation 
\bea 
  \widehat\Phi_{a_k,b_k}[{\vec x_k}] = \Phi_{a_k,x_{k,1}}\cdot  
  \Phi_{x_{k,1},x_{k,2}} \cdot \Phi_{x_{k,2},x_{k,3}}\cdot \ldots 
  \cdot \Phi_{x_{k,n-1},x_{k,n}}  \cdot \Phi_{x_{k,n(k)},b_k} 
\eea 
for the chain-product of open string vertex operators. 
Here we define $\widehat\Phi_{a_k,b_k}[\vec 0]=\Phi_{a_k,b_k}$.

To extract the superpotential, one 
can probe it by evaluating an appropriate matter field 
correlator in the instanton background. 
The CFT allows one to compute it in physical normalisation 
which combines the superpotential part $Y$ with 
the matter field K\"ahler metrics like 
\bea 
\label{correla} 
     && \langle \Phi_{a_1,b_1}\cdot\ldots\cdot   \Phi_{a_M,b_M}  
\rangle_{E2-{\rm inst}} =  
{e^{{\cal K}\over 2}\, Y_{\Phi_{a_1,b_1},\ldots, \Phi_{a_M,b_M}} \over \sqrt{ K_{a_1,b_1}\cdot\ldots\cdot K_{a_M,b_M} 
  }}. 
\eea 
\vskip 0.3cm 
 
In \cite{Blumenhagen:2006xt} a general expression for the single E2-instanton contribution 
to the charged matter superpotential was proposed involving    
the evaluation of  the following zero mode integral over disc and one-loop 
open string CFT amplitudes 
\bea 
\label{mainall} 
     &&\hskip -1cm \langle \Phi_{a_1,b_1}\cdot\ldots\cdot   \Phi_{a_M,b_M}  
\rangle_{E2}  
 = {V_3\over g_s}\,\int d^4 x\, d^2\theta \,\, 
       \sum_{\rm conf.}\,\,  {\textstyle  
  \prod_{a} \bigl(\prod_{i=1}^{ [\Xi\cap 
             \Pi_a]^+}  d\lambda_a^i\bigr)\, 
               \bigl( \prod_{i=1}^{ [\Xi\cap 
             \Pi_a]^-}  d\ov{\lambda}_a^i\bigr) } \nonumber \\ 
   && \phantom{a} \exp ({-S_{E2}}) \,\, 
           \exp \left({Z'_0(E2)}\right) \,\, 
\langle \widehat\Phi_{a_1,b_1}[\vec x_1]   \rangle_{\lambda_{a_1},\ov{\lambda}_{b_1}}\cdot 
            \ldots \cdot  \langle \widehat\Phi_{a_L,b_L}[\vec x_L] 
          \rangle_{\lambda_{a_L},\ov{\lambda}_{b_L}} \, . 
\eea 
 
\begin{figure}[h] 
\begin{center} 
 \includegraphics[width=0.7\textwidth]{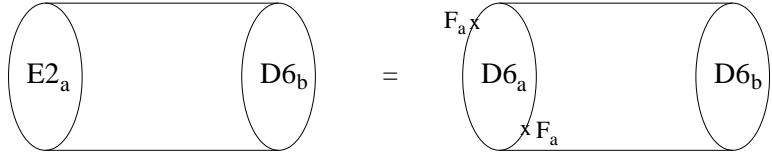} 
\end{center} 
\caption{\small Relation between instantonic one-loop amplitudes and corresponding 
 gauge  threshold corrections}\label{nicerel} 
\end{figure}

\begin{figure}[h] 
\begin{center} 
 \includegraphics[width=0.7\textwidth]{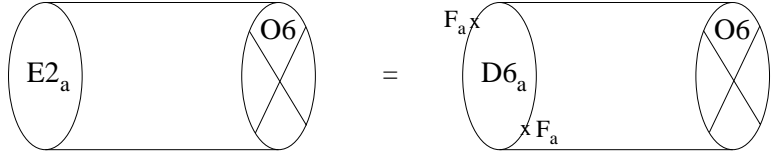} 
\end{center} 
\caption{\small Relation between instantonic M\"obius amplitude  and corresponding 
 gauge  threshold corrections}\label{nicerelm} 
\end{figure}

\vskip 0.3cm 
For simplicity, we do not consider the case that matter fields 
are also assigned to string loop diagrams. 
The one-loop contributions are annulus diagrams for open strings 
with one boundary on the E2-instanton and the other  
boundary on the various D6-branes and M\"obius diagrams with boundary on the E2-instanton 
\bea 
\label{loop_diagram} 
   \langle 1 \rangle^{\text{1-loop}}=   
    Z'_0(E2) = 
   {\textstyle \sum_b  {Z'}^A ({\rm E2}_a,{\rm D6}_b)  
    +  {Z'}^M({\rm E2}_a,{\rm O6})}\;.  
\eea 
Here $Z'$ means that we only sum over the massive open 
string states in the loop amplitude, as the zero modes 
are taken care of explicitly.
It was shown that these instantonic open string loop diagrams 
are identical to the one-loop threshold corrections  
$T^A({\rm D6}_a,{\rm D6}_b)$. 
Diagrammatically we have the intriguing 
relation shown in figure \ref{nicerel} and in figure \ref{nicerelm}, 
which holds for the even spin structures\footnote{The contribution of 
the CP-odd R$^-$ sector is expected to yield corrections 
to the $\theta$-angle.}. 
 
\noindent 
The annulus threshold corrections can be computed, leading to 
\bea 
\label{thresholdannu} 
  Z^A({\rm E2}_a,{\rm D6}_b)=\int_0^\infty  {dt\over t}\,  
  \sum_{\alpha,\beta\neq(\frac{1}{2},\frac{1}{2})} (-1)^{2(\alpha+\beta)}\,  
  {\vartheta''\thba{\alpha}{\beta}(it) \over\eta^3(it)}\,\,  
  A^{\rm CY}_{ab}\thba{\alpha}{\beta}(it) \; 
\eea 
and the M\"obius strip amplitude for the instanton,  
which as we explained must be   
invariant under the orientifold projection, yields 
\bea 
\label{thresholdmoe} 
  Z^M({\rm E2}_a,{\rm O6})=\pm \int_0^\infty  {dt\over t}\,  
  \sum_{\alpha,\beta\neq(\frac{1}{2},\frac{1}{2})} (-1)^{2(\alpha+\beta)}\,  
  {\vartheta''\thba{\alpha}{\beta}\left(it+{1\over 2}\right) \over\eta^3 \left(it+{1\over 2}\right)}\,\,  
  A^{\rm CY}_{aa}\thba{\alpha}{\beta}\left(it+{\textstyle {1\over 2}}\right) \;. 
\eea 
The overall plus sign is for $O(1)$ instantons, reflecting the fact 
that only for these the $x_\mu$ and $\theta_\alpha$ zero modes 
survive the orientifold projection. Note that up to the argument,  
the M\"obius thresholds are  $Z^A({\rm E2}_a,{\rm D6}_a)$. 
Therefore, for rigid branes the massless sector reflects the 
number of four  bosonic and two fermionic zero modes.  
In section \ref{sectionfinst} we will discuss the number of zero modes  
if  $b_1(\Xi)>0$. 
All these stringy threshold corrections are known to be non-holomorphic. 
Therefore, it is not immediately obvious in which sense 
the expression (\ref{mainall}) is meant and how one can extract the 
holomorphic superpotential part $Y$ from it.  
 
The CFT disc amplitudes in (\ref{mainall}) are also not holomorphic but combine 
non-holomorphic K\"ahler potential contributions and 
holomorphic superpotential contributions in the usual way \cite{Cremades:2003qj,Cvetic:2003ch,Abel:2003vv,Cremades:2004wa}: 
\bea 
\label{disci} 
\langle \widehat\Phi_{a,b}[\vec x]   \rangle_{\lambda_{a},\ov{\lambda}_{b}}&=& 
{e^{{\cal K}\over 2}\, Y_{\lambda_a \Phi_{a,x_1}\Phi_{x_1,x_2}\ldots 
   \Phi_{x_N,b}\,  \ov\lambda_b}  
    \over \sqrt{ K_{\lambda_a,a} \, K_{a,x_1}\, \ldots  K_{x_n,b}\,  
          K_{b,\ov\lambda_b} }}\\ 
   &=& {e^{{\cal K}\over 2}\, Y_{\lambda_a\, \widehat\Phi_{a,b}[x]\, \ov\lambda_b}  
    \over \sqrt{ K_{\lambda_a,a} \, \widehat K_{a,b}[x]\,  
          K_{b,\ov\lambda_b} }}\, . 
\eea 
Due to the Kaplunovsky-Louis formula (\ref{gceffFT}),  
the stringy one-loop amplitudes are known to include  
the holomorphic Wilsonian part and contributions from wave-function 
normalisation. Applied to the instanton one-loop amplitudes appearing 
in $Z_0(E2_a)$, we write 
\bea 
\label{runn} 
Z_0(E2_a)&=&   
-8\pi^2\,\Re ( {f^{(1)}_a}) - {b_a\over 2}\ln\left( {M_{p}^2\over 
    \mu^2}\right) - {c_a\over 2}\, {\cal K}_{\rm tree}\\  
&&\phantom{aaa}  -  
  \, \ln\left( {V_3\over g_s}\right)_{\rm tree} + \sum_b {|I_{ab} N_b|\over 2} 
\ln\left[\det K^{ab} \right]_{\rm tree}\, , \nonumber 
\eea 
where for the brane and instanton configuration in question the coefficients 
are   
\bea 
   b_a=\sum_b {|I_{ab} N_b|\over 2} -3,\quad\quad  
   c_a=\sum_b {|I_{ab} N_b|\over 2} -1. 
\eea 
The constant contributions arise from the M\"obius 
amplitude. 
Inserting (\ref{disci}) and (\ref{runn}) in (\ref{mainall}), 
one realises that the K\"ahler metrics involving an instanton 
zero mode and a matter field precisely cancel out, so that 
only the matter metrics survive, as required by the general form 
(\ref{correla}). Moreover, the term $\exp({\cal K}/2)$ comes 
out just right due to the rule that each disc 
contains precisely two instanton zero modes. 
The holomorphic piece in (\ref{correla}) 
can therefore be expressed entirely in terms of other 
holomorphic quantities like holomorphic Yukawa couplings, 
the holomorphic instanton action and the one-loop 
holomorphic Wilsonian gauge kinetic function 
on the E2-brane: 
\bea 
\label{final} 
    Y_{\Phi_{a_1,b_1},\ldots, \Phi_{a_M,b_M}}&=& 
       \sum_{\rm conf.}\,\,  \sign_{\rm conf}\, 
  \,\, \exp ({-S_{E2}})_{\rm tree} \, \, 
        \, \exp \left(-f^{(1)}_a\right)  
\,  \nonumber \\ 
&&\phantom{aaaaa} Y_{\lambda_{a_1}\, \widehat\Phi_{a_1,b_1}[\vec x_1]\, \ov\lambda_{b_1}} 
\cdot\ldots\cdot Y_{\lambda_{a_1}\, \widehat\Phi_{a_L,b_L}[\vec x_L]\, 
         \ov\lambda_{b_L}} . 
\eea 
This explicitly shows that knowing the tree-level K\"ahler potentials, 
computing  the matter field correlator in the instanton background 
up to one-loop level in $g_s$ is sufficient to deduce the 
Wilsonian holomorphic instanton generated superpotential. 
Higher order  corrections in $g_s$ only come from loop corrections 
to the K\"ahler potentials.

\subsection{Instanton corrections to the gauge kinetic functions} 
\label{sectionfinst} 
 
So far we have discussed space-time instanton corrections 
to the superpotential. These involved one-loop determinants, 
which are given by annulus vacuum diagrams with at least 
one E2 boundary. These are related to one-loop gauge threshold 
corrections to the gauge theory on a D6-brane wrapping 
the same cycle as the E2 instanton. 
 
Now we can ask what other corrections these space-time 
instantons can induce. By applying S- and T-dualities to 
the story of world-sheet instanton corrections 
in the heterotic string, we expect that there can also 
be {\it E2-instanton corrections} to the holomorphic  
gauge kinetic functions. 
In the heterotic case, similar to the topological Type II 
string,  such corrections arise from string world-sheets 
of Euler characteristic zero, i.e. here from world-sheets 
with two boundaries.  
Therefore, we expect such corrections to appear for E2-instantons 
admitting one complex open string modulus, i.e. 
those wrapping a  three-cycle with Betti number  $b_1(\Xi)=1$.   
 
Let us start by discussing the instanton zero mode structure 
for such a cycle. 
First let us provide the form of the vertex operators. 
 The bosonic fields in the $(-1)$ ghost 
picture are 
\bea 
      V^{(-1)}_{y}(z)=y\  e^{-\varphi (z)}\, \Sigma_{h={1\over 2},q=\pm 1}(z) 
\eea 
which, before the orientifold projection, are accompanied 
by the two pairs of fermionic zero modes 
\bea 
 V^{(-1/2)}_{\mu}(z)=\mu_\alpha\  e^{-{\varphi(z)\over 2}}\, S^\alpha(z)\, \, 
      \Sigma_{h={3\over 8},q=-{1\over 2}}(z)\, 
\eea 
and 
\bea 
 V^{(-1/2)}_{\ov\mu}(z)=\ov\mu_{\dot\alpha}\   e^{-{\varphi(z)\over 2}}\,{S}^{\dot\alpha}(z)\,\,  
      \Sigma_{h={3\over 8},q=+{1\over 2}}(z)\,. 
\eea 
Now one has to distinguish two cases depending on how the  
anti-holomorphic involution $\ov\sigma$ acts on the open string 
modulus $Y$ 
\bea 
      \ov\sigma:y\to \pm y. 
\eea 
In the case that $y$ is invariant under $\ov\sigma$, called first kind in the 
following,  the orientifold 
projection acts in the same way as for the 4D fields $X_\mu$, i.e. 
the two bosonic zero modes $y$ and the two fermionic zero modes 
$\ov\mu$ survive. In the other case, dubbed second kind,  the bosonic zero mode is projected 
out and only the fermionic modulino zero mode $\mu$ survives\footnote{By duality, this 
distinction is related to the two kinds of deformations 
of genus $g$ curves studied in \cite{Beasley:2005iu}. The first kind are the 
curves moving in families, i.e. transversal deformation of the curve. 
The second kind is related to  
the deformations coming with the genus $g$ of the curve.}. 
Therefore, in the absence of any additional zero modes, for instance 
from E2-D6 intersections, the zero mode measure in any instanton 
amplitude assumes the following 
form 
\bea 
      \int d^4 x\, d^2\theta\, d^2 y\, d^2 \ov\mu\ e^{-S_{E2}}\ldots,\quad\quad {\rm for}\  
         \ov\sigma:y\to y 
\eea 
and 
\bea 
      \int d^4 x\, d^2\theta\, d^2 \mu \ e^{-S_{E2}}\ldots,\quad\quad {\rm for}\  
         \ov\sigma:y\to -y. 
\eea 
 
\begin{figure}[h] 
\begin{center} 
 \includegraphics[width=0.85\textwidth]{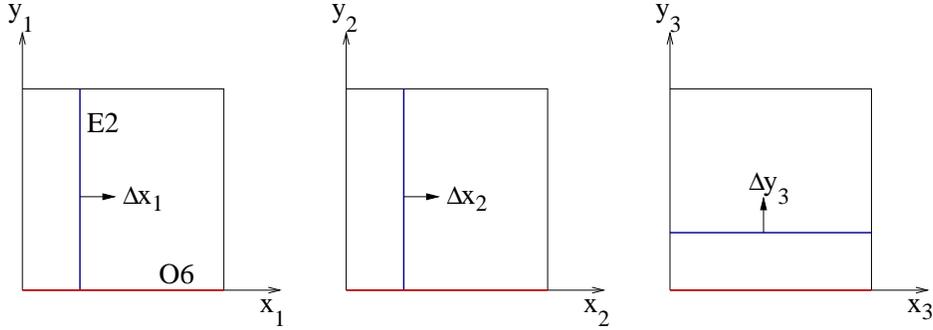} 
\end{center} 
\caption{\small Deformations of an instanton which is invariant 
under the orientifold projection}\label{configu} 
\end{figure}

\noindent 
As an example consider the set-up in figure \ref{configu}  with $\ov\sigma: y_i\to -y_i$. 
Here the deformations $\Delta x_{1,2}$ are of the first kind and 
$\Delta y_3$ is of the second kind. 
 
\begin{figure}[h] 
\begin{center} 
 \includegraphics[width=0.5\textwidth]{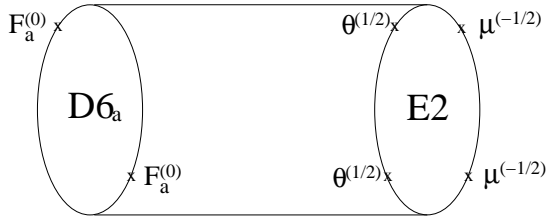} 
\end{center} 
\caption{Annulus diagram for E2-instanton correction to $f_a$. 
The upper indices give the ghost number of the vertex operators.}\label{finst} 
\end{figure}

Now, it is clear that an instanton with precisely one set of  fermionic zero 
modes of the second  kind and no additional zero modes 
can generate a correction to the $SU(N_a)$ gauge kinetic function. The  
instanton amplitude takes on the following form 
\bea 
\label{gaugekin} 
     \langle F_a(p_1)\, F_a(p_2)  
\rangle_{E2}  
 = \int d^4 x\, d^2\theta \, d^2 \mu\ 
       \exp ({-S_{E2}}) \, 
         \, \exp \left({Z'_0 (E2)}\right)\,\,  A_{F_a^2}(E2,D6_a) \nonumber 
\eea 
where $A_{F_a^2}(E2,D6_a)$ is the annulus diagram in figure \ref{finst}, which 
absorbs all the appearing fermionic zero modes 
and where the gauge boson vertex operators in the $(0)$-ghost picture have 
the usual form 
\bea 
   V^{(0)}_{A}(z)=\epsilon^\mu\  (\partial_\mu  X(z) + i(p\cdot \psi)\, \psi_\mu(z)) \, e^{i p\cdot X(z)}. 
\eea 
 
Analogous to  world-sheet instantons, these diagrams can be generalised 
to multi ${\rm tr}(W^2)^h$ couplings. Just from the zero mode counting 
one immediately sees that they can be generated by E2-instantons 
with $h$ sets of complex deformation zero modes of the second kind and no 
other additional zero modes. Then, besides the annulus diagram in figure \ref{finst}, 
there are $h-1$ similar diagrams. On the D6$_a$ brane one inserts  
two gauginos  in the $(+1/2)$ ghost picture  and on the E2 boundary two $\mu$ modulinos 
in the $(-1/2)$ ghost picture.  
Clearly, once the internal ${\cal N}=2$ superconformal field theory is known, as 
for toroidal orbifolds or Gepner models, these annulus diagrams 
can be computed explicitly. They involve up to four-point functions 
of vertex operators on an annulus  world-sheet with the two boundaries 
on the E2 and the D6$_a$ brane. 
Very similar to the ${\cal N}=2$ open string sectors for loop-corrections 
to $f_a$, one expects these instanton diagrams to also contain  a sum over 
world-sheet instantons. Therefore, the generic E2-instanton contribution 
to the holomorphic gauge kinetic functions has the moduli dependence 
$f^{\rm np}\bigl(e^{-U^c_I}, e^{-T^c_i} \bigr)$.

\subsection{Instanton corrections to the FI-terms} 
\label{sectioninstfi} 
 
Having shown that E2-instanton corrections to the 
gauge couplings are possible, it is natural to 
investigate whether such instantons also contribute 
to the FI-terms for the $U(1)$ gauge symmetries on the 
D6-branes. As we have seen in section \ref{secfit}, 
at the one-loop level the contributions to 
the gauge couplings and to the FI-terms have the 
same functional form.  
 
\begin{figure}[h] 
\begin{center} 
 \includegraphics[width=0.5\textwidth]{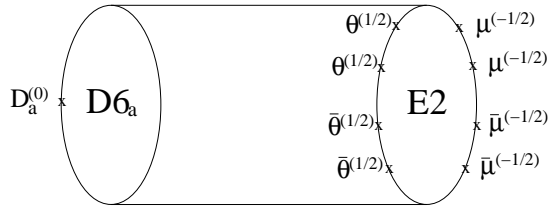} 
\end{center} 
\caption{Annulus diagram for E2-instanton correction to $\xi_a$. 
The upper indices give the ghost number of the vertex operators.}\label{dinst} 
\end{figure}

Assume now that, as in the last section, in the background with intersecting 
D6-branes we can find an E2-instanton with only two $\theta$ fermionic zero 
modes and two additional fermionic zero modes related to a deformation 
of the E2.  
If now similar to the D6-branes we could break supersymmetry 
on the E2-branes by a slight deformation of the complex 
structure, then we would expect four $\theta$-like, 
four $\mu$-like  and two $y$-like zero modes. As shown in figure \ref{dinst}, 
these could generate an FI-term on the D6-branes.  
However, since the E2-brane must be  invariant, i.e. an $O(1)$ instanton, 
 under the orientifold projection, a complex structure deformation 
does not necessarily break supersymmetry on the E2-instanton. In this case, the 
analogous situation to the one-loop D6-brane generation of 
the FI-term cannot happen. 
 
However, there is another mechanism to generate an FI-term on 
the E2-instanton, namely by turning on the $\int_{\Xi} C_3$ 
modulus through the three-cycle the E2-instanton is wrapping. 
This also appears in the (generalised) calibration condition \cite{Gutowski:1999tu,Gmeiner:2006ni} 
for supersymmetry on the E2-brane. 
Therefore, it is possible that the one-loop diagram 
in figure \ref{dinst} indeed generates an FI-term  
on the D6$_a$ brane once the $C_3$ flux through 
the E2 is non-zero. 
 
Here we will leave a further study of the concrete 
instanton amplitudes for $g^{-2}_a$ and $\xi_a$ and their relation for future work and conclude 
that just from fermionic zero mode counting, 
we have evidence that E2-instanton corrections 
to both the gauge kinetic functions and 
the FI-terms are likely to appear. 
 
\section{Conclusions} 
 
In this paper we have investigated a number of aspects 
related to loop and D-brane instanton corrections 
to intersecting D6-brane models in Type IIA 
orientifolds. In particular, we have revisited the  
computation of one-loop corrections 
to the FI-terms. 
 
Using results for the gauge threshold corrections in intersecting D6-brane models 
on a toroidal orientifold,  we explicitly computed  the Wilsonian 
holomorphic gauge coupling in this setup. On the way, exploiting  holomorphy 
and the Shifman-Vainshtein, respectively Kaplunovsky-Louis 
formula, it was possible  
to constrain the form of the matter field K\"ahler 
metrics. 
In the second part, we discussed E2-brane instanton 
corrections to the superpotential, the 
gauge kinetic function and the FI-terms.  
For the first, we showed in which sense one can 
extract the form of the holomorphic superpotential 
from a superconformal field theory correlation function of matter fields 
in the E2-instanton background. 
 
Moreover, we showed that E2-instantons wrapping  
a three-cycle which has precisely one complex  deformation 
and no matter zero modes,  
can in principle contribute to the gauge kinetic function 
for a gauge theory on a stack of D6-branes. 
By turning on the R-R three-form modulus, also instanton 
corrections to the FI-terms become possible. 
A more detailed investigation of the appearing 
annulus diagrams is necessary to eventually 
establish the appearance of these instanton 
corrections, but our first steps indicate 
that such corrections are indeed present in  
${\cal N}=1$ D-brane vacua.

%%%%%%%%%%%%%%%%%%%%%%%%%%%%%%%%%%%%%%%%%%%%%%% 
%%%%%%%%%%%%%%%%%%%%%%%%%%%%%%%%%%%%%%%%%%%%%%% 
%%%%%%%%%%%%%%%%%%%%%%%%%%%%%%%%%%%%%%%%%%%%%%% 
%%%%%%%%%%%%%%%%%%%%%%%%%%%%%%%%%%%%%%%%%%%%%%% 

\vskip 1cm 
 {\noindent  {\Large \bf Acknowledgements}} 
 \vskip 0.5cm 
We would like to thank Emilian Dudas, Michael Haack, Sebastian Moster,  
Erik Plauschinn, Stephan Stieberger, Angel Uranga  and 
Timo Weigand for interesting discussions. 
This work is supported in part by the European Community's Human 
Potential Programme under contract MRTN-CT-2004-005104 `Constituents, 
fundamental forces and symmetries of the universe'. 
 
%%%%%%%%%%%%%%%%%%%%%%%%%%%%%%%%%%%%%%%%%%%%%%% 
%%%%%%%%%%%%%%%%%%%%%%%%%%%%%%%%%%%%%%%%%%%%%%% 
%%%%%%%%%%%%%%%%%%%%%%%%%%%%%%%%%%%%%%%%%%%%%%% 
%%%%%%%%%%%%%%%%%%%%%%%%%%%%%%%%%%%%%%%%%%%%%%% 
%%%%%%%%%%%%%%%%%%%%%%%%%%%%%%%%%%%%%%%%%%%%%%% 
%%%%%%%%%%%%%%%%%%%%%%%%%%%%%%%%%%%%%%%%%%%%%%% 
%%%%%%%%%%%%%%%%%%%%%%%%%%%%%%%%%%%%%%%%%%%%%%% 
%%%%%%%%%%%%%%%%%%%%%%%%%%%%%%%%%%%%%%%%%%%%%%% 

\clearpage 
\nocite{*} 
\bibliography{rev} 

\providecommand{\href}[2]{#2}\begingroup\raggedright\begin{thebibliography}{10}

\bibitem{Uranga:2003pz}
A.~M. Uranga, ``Chiral four-dimensional string compactifications with
  intersecting D-branes,'' {\em Class. Quant. Grav.} {\bf 20} (2003)
  S373--S394,
\href{http://www.arXiv.org/abs/hep-th/0301032}{{\tt hep-th/0301032}}.
%%CITATION = HEP-TH/0301032;%%.

\bibitem{Lust:2004ks}
D.~L{\"u}st, ``Intersecting brane worlds: A path to the standard model?,'' {\em
  Class. Quant. Grav.} {\bf 21} (2004) S1399--1424,
\href{http://www.arXiv.org/abs/hep-th/0401156}{{\tt hep-th/0401156}}.
%%CITATION = HEP-TH/0401156;%%.

\bibitem{Blumenhagen:2005mu}
R.~Blumenhagen, M.~Cvetic, P.~Langacker, and G.~Shiu, ``Toward realistic
  intersecting D-brane models,'' {\em Ann. Rev. Nucl. Part. Sci.} {\bf 55}
  (2005) 71--139,
\href{http://www.arXiv.org/abs/hep-th/0502005}{{\tt hep-th/0502005}}.
%%CITATION = HEP-TH/0502005;%%.

\bibitem{Blumenhagen:2006ci}
R.~Blumenhagen, B.~K{\"o}rs, D.~L{\"u}st, and S.~Stieberger, ``Four-dimensional
  string compactifications with D-branes, orientifolds and fluxes,''
\href{http://www.arXiv.org/abs/hep-th/0610327}{{\tt hep-th/0610327}}.
%%CITATION = HEP-TH 0610327;%%.

\bibitem{Lust:2003ky}
D.~L{\"u}st and S.~Stieberger, ``Gauge threshold corrections in intersecting
  brane world models,''
\href{http://www.arXiv.org/abs/hep-th/0302221}{{\tt hep-th/0302221}}.
%%CITATION = HEP-TH 0302221;%%.

\bibitem{letter}
N.~Akerblom, R.~Blumenhagen, D.~L{\"u}st, and M.~Schmidt-Sommerfeld,
  ``Thresholds for Intersecting D-branes Revisited,''
\href{http://www.arXiv.org/abs/arXiv:0705.2150 [hep-th]}{{\tt arXiv:0705.2150
  [hep-th]}}.
%%CITATION = ARXIV:0705.2150;%%.

\bibitem{Lawrence:2004sm}
A.~Lawrence and J.~McGreevy, ``D-terms and D-strings in open string models,''
  {\em JHEP} {\bf 10} (2004) 056,
\href{http://www.arXiv.org/abs/hep-th/0409284}{{\tt hep-th/0409284}}.
%%CITATION = HEP-TH/0409284;%%.

\bibitem{Derendinger:1991hq}
J.~P. Derendinger, S.~Ferrara, C.~Kounnas, and F.~Zwirner, ``On loop
  corrections to string effective field theories: Field dependent gauge
  couplings and sigma model anomalies,'' {\em Nucl. Phys.} {\bf B372} (1992)
145--188.
%%CITATION = NUPHA,B372,145;%%.

\bibitem{Derendinger:1991kr}
J.-P. Derendinger, S.~Ferrara, C.~Kounnas, and F.~Zwirner, ``All loop gauge
  couplings from anomaly cancellation in string effective theories,'' {\em
  Phys. Lett.} {\bf B271} (1991)
307--313.
%%CITATION = PHLTA,B271,307;%%.

\bibitem{LopesCardoso:1991zt}
G.~Lopes~Cardoso and B.~A. Ovrut, ``A Green-Schwarz mechanism for D = 4, N=1
  supergravity anomalies,'' {\em Nucl. Phys.} {\bf B369} (1992)
351--372.
%%CITATION = NUPHA,B369,351;%%.

\bibitem{Ibanez:1992hc}
L.~E. Ibanez and D.~L{\"u}st, ``Duality anomaly cancellation, minimal string
  unification and the effective low-energy Lagrangian of 4-D strings,'' {\em
  Nucl. Phys.} {\bf B382} (1992) 305--364,
\href{http://www.arXiv.org/abs/hep-th/9202046}{{\tt hep-th/9202046}}.
%%CITATION = HEP-TH 9202046;%%.

\bibitem{Kaplunovsky:1994fg}
V.~Kaplunovsky and J.~Louis, ``Field dependent gauge couplings in locally
  supersymmetric effective quantum field theories,'' {\em Nucl. Phys.} {\bf
  B422} (1994) 57--124,
\href{http://www.arXiv.org/abs/hep-th/9402005}{{\tt hep-th/9402005}}.
%%CITATION = HEP-TH 9402005;%%.

\bibitem{Kaplunovsky:1995jw}
V.~Kaplunovsky and J.~Louis, ``On Gauge couplings in string theory,'' {\em
  Nucl. Phys.} {\bf B444} (1995) 191--244,
\href{http://www.arXiv.org/abs/hep-th/9502077}{{\tt hep-th/9502077}}.
%%CITATION = HEP-TH 9502077;%%.

\bibitem{Blumenhagen:2006xt}
R.~Blumenhagen, M.~Cvetic, and T.~Weigand, ``Spacetime instanton corrections in
  4D string vacua - the seesaw mechanism for D-brane models,'' {\em Nucl.
  Phys.} {\bf B771} (2007) 113--142,
\href{http://www.arXiv.org/abs/hep-th/0609191}{{\tt hep-th/0609191}}.
%%CITATION = HEP-TH/0609191;%%.

\bibitem{Ibanez:2006da}
L.~E. Ibanez and A.~M. Uranga, ``Neutrino Majorana masses from string theory
  instanton effects,'' {\em JHEP} {\bf 03} (2007) 052,
\href{http://www.arXiv.org/abs/hep-th/0609213}{{\tt hep-th/0609213}}.
%%CITATION = HEP-TH/0609213;%%.

\bibitem{Haack:2006cy}
M.~Haack, D.~Krefl, D.~L{\"u}st, A.~Van~Proeyen, and M.~Zagermann, ``Gaugino
  condensates and D-terms from D7-branes,'' {\em JHEP} {\bf 01} (2007) 078,
\href{http://www.arXiv.org/abs/hep-th/0609211}{{\tt hep-th/0609211}}.
%%CITATION = HEP-TH/0609211;%%.

\bibitem{Abel:2006yk}
S.~A. Abel and M.~D. Goodsell, ``Realistic Yukawa couplings through instantons
  in intersecting brane worlds,''
\href{http://www.arXiv.org/abs/hep-th/0612110}{{\tt hep-th/0612110}}.
%%CITATION = HEP-TH/0612110;%%.

\bibitem{Bianchi:2007wy}
M.~Bianchi, F.~Fucito, and J.~F. Morales, ``D-brane Instantons on the T6/Z3
  orientifold,''
\href{http://www.arXiv.org/abs/arXiv:0704.0784 [hep-th]}{{\tt arXiv:0704.0784
  [hep-th]}}.
%%CITATION = ARXIV:0704.0784;%%.

\bibitem{Cvetic:2007ku}
M.~Cvetic, R.~Richter, and T.~Weigand, ``Computation of D-brane instanton
  induced superpotential couplings: Majorana masses from string theory,''
\href{http://www.arXiv.org/abs/hep-th/0703028}{{\tt hep-th/0703028}}.
%%CITATION = HEP-TH/0703028;%%.

\bibitem{Ibanez:2007rs}
L.~E. Ibanez, A.~N. Schellekens, and A.~M. Uranga, ``Instanton Induced Neutrino
  Majorana Masses in CFT Orientifolds with MSSM-like spectra,''
\href{http://www.arXiv.org/abs/arXiv:0704.1079 [hep-th]}{{\tt arXiv:0704.1079
  [hep-th]}}.
%%CITATION = ARXIV:0704.1079;%%.

\bibitem{Billo:2002hm}
M.~Billo {\em et al.}, ``Classical gauge instantons from open strings,'' {\em
  JHEP} {\bf 02} (2003) 045,
\href{http://www.arXiv.org/abs/hep-th/0211250}{{\tt hep-th/0211250}}.
%%CITATION = HEP-TH 0211250;%%.

\bibitem{Florea:2006si}
B.~Florea, S.~Kachru, J.~McGreevy, and N.~Saulina, ``Stringy instantons and
  quiver gauge theories,''
\href{http://www.arXiv.org/abs/hep-th/0610003}{{\tt hep-th/0610003}}.
%%CITATION = HEP-TH 0610003;%%.

\bibitem{Akerblom:2006hx}
N.~Akerblom, R.~Blumenhagen, D.~L{\"u}st, E.~Plauschinn, and
  M.~Schmidt-Sommerfeld, ``Non-perturbative SQCD Superpotentials from String
  Instantons,''
\href{http://www.arXiv.org/abs/hep-th/0612132}{{\tt hep-th/0612132}}.
%%CITATION = HEP-TH/0612132;%%.

\bibitem{Argurio:2007vq}
R.~Argurio, M.~Bertolini, G.~Ferretti, A.~Lerda, and C.~Petersson, ``Stringy
  Instantons at Orbifold Singularities,''
\href{http://www.arXiv.org/abs/arXiv:0704.0262 [hep-th]}{{\tt arXiv:0704.0262
  [hep-th]}}.
%%CITATION = ARXIV:0704.0262;%%.

\bibitem{Bianchi:2007fx}
M.~Bianchi and E.~Kiritsis, ``Non-perturbative and Flux superpotentials for
  Type I strings on the Z3 orbifold,''
\href{http://www.arXiv.org/abs/hep-th/0702015}{{\tt hep-th/0702015}}.
%%CITATION = HEP-TH/0702015;%%.

\bibitem{Bianchi:2007ft}
M.~Bianchi, S.~Kovacs, and G.~Rossi, ``Instantons and supersymmetry,''
\href{http://www.arXiv.org/abs/hep-th/0703142}{{\tt hep-th/0703142}}.
%%CITATION = HEP-TH/0703142;%%.

\bibitem{Witten:1999eg}
E.~Witten, ``World-sheet corrections via D-instantons,'' {\em JHEP} {\bf 02}
  (2000) 030,
\href{http://www.arXiv.org/abs/hep-th/9907041}{{\tt hep-th/9907041}}.
%%CITATION = HEP-TH 9907041;%%.

\bibitem{Green:1997tv}
M.~B. Green and M.~Gutperle, ``Effects of D-instantons,'' {\em Nucl. Phys.}
  {\bf B498} (1997) 195--227,
\href{http://www.arXiv.org/abs/hep-th/9701093}{{\tt hep-th/9701093}}.
%%CITATION = HEP-TH 9701093;%%.

\bibitem{Gutperle:1997iy}
M.~Gutperle, ``Aspects of D-instantons,''
\href{http://www.arXiv.org/abs/hep-th/9712156}{{\tt hep-th/9712156}}.
%%CITATION = HEP-TH 9712156;%%.

\bibitem{Anastasopoulos:2006hn}
P.~Anastasopoulos, M.~Bianchi, G.~Sarkissian, and Y.~S. Stanev, ``On gauge
  couplings and thresholds in type I gepner models and otherwise,'' {\em JHEP}
  {\bf 03} (2007) 059,
\href{http://www.arXiv.org/abs/hep-th/0612234}{{\tt hep-th/0612234}}.
%%CITATION = HEP-TH/0612234;%%.

\bibitem{Poppitz:1998dj}
E.~Poppitz, ``On the one loop Fayet-Iliopoulos term in chiral four dimensional
  type I orbifolds,'' {\em Nucl. Phys.} {\bf B542} (1999) 31--44,
\href{http://www.arXiv.org/abs/hep-th/9810010}{{\tt hep-th/9810010}}.
%%CITATION = HEP-TH/9810010;%%.

\bibitem{Bain:2000fb}
P.~Bain and M.~Berg, ``Effective action of matter fields in four-dimensional
  string orientifolds,'' {\em JHEP} {\bf 04} (2000) 013,
\href{http://www.arXiv.org/abs/hep-th/0003185}{{\tt hep-th/0003185}}.
%%CITATION = HEP-TH/0003185;%%.

\bibitem{Shifman:1986zi}
M.~A. Shifman and A.~I. Vainshtein, ``Solution of the anomaly puzzle in susy
  gauge theories and the Wilson operator expansion,'' {\em Nucl. Phys.} {\bf
  B277} (1986)
456.
%%CITATION = NUPHA,B277,456;%%.

\bibitem{Lust:2004cx}
D.~L{\"u}st, P.~Mayr, R.~Richter, and S.~Stieberger, ``Scattering of gauge,
  matter, and moduli fields from intersecting branes,'' {\em Nucl. Phys.} {\bf
  B696} (2004) 205--250,
\href{http://www.arXiv.org/abs/hep-th/0404134}{{\tt hep-th/0404134}}.
%%CITATION = HEP-TH 0404134;%%.

\bibitem{Kors:2003wf}
B.~K{\"o}rs and P.~Nath, ``Effective action and soft supersymmetry breaking for
  intersecting D-brane models,'' {\em Nucl. Phys.} {\bf B681} (2004) 77--119,
\href{http://www.arXiv.org/abs/hep-th/0309167}{{\tt hep-th/0309167}}.
%%CITATION = HEP-TH/0309167;%%.

\bibitem{Antoniadis:1999ge}
I.~Antoniadis, C.~Bachas, and E.~Dudas, ``Gauge couplings in four-dimensional
  type I string orbifolds,'' {\em Nucl. Phys.} {\bf B560} (1999) 93--134,
\href{http://www.arXiv.org/abs/hep-th/9906039}{{\tt hep-th/9906039}}.
%%CITATION = HEP-TH 9906039;%%.

\bibitem{Berg:2004ek}
M.~Berg, M.~Haack, and B.~Kors, ``Loop corrections to volume moduli and
  inflation in string theory,'' {\em Phys. Rev.} {\bf D71} (2005) 026005,
\href{http://www.arXiv.org/abs/hep-th/0404087}{{\tt hep-th/0404087}}.
%%CITATION = HEP-TH/0404087;%%.

\bibitem{Cremades:2003qj}
D.~Cremades, L.~E. Ibanez, and F.~Marchesano, ``Yukawa couplings in
  intersecting D-brane models,'' {\em JHEP} {\bf 07} (2003) 038,
\href{http://www.arXiv.org/abs/hep-th/0302105}{{\tt hep-th/0302105}}.
%%CITATION = HEP-TH 0302105;%%.

\bibitem{Cvetic:2003ch}
M.~Cvetic and I.~Papadimitriou, ``Conformal field theory couplings for
  intersecting D-branes on orientifolds,'' {\em Phys. Rev.} {\bf D68} (2003)
  046001,
\href{http://www.arXiv.org/abs/hep-th/0303083}{{\tt hep-th/0303083}}.
%%CITATION = HEP-TH 0303083;%%.

\bibitem{Abel:2003vv}
S.~A. Abel and A.~W. Owen, ``Interactions in intersecting brane models,'' {\em
  Nucl. Phys.} {\bf B663} (2003) 197--214,
\href{http://www.arXiv.org/abs/hep-th/0303124}{{\tt hep-th/0303124}}.
%%CITATION = HEP-TH/0303124;%%.

\bibitem{Cremades:2004wa}
D.~Cremades, L.~E. Ibanez, and F.~Marchesano, ``Computing Yukawa couplings from
  magnetized extra dimensions,'' {\em JHEP} {\bf 05} (2004) 079,
\href{http://www.arXiv.org/abs/hep-th/0404229}{{\tt hep-th/0404229}}.
%%CITATION = HEP-TH/0404229;%%.

\bibitem{Beasley:2005iu}
C.~Beasley and E.~Witten, ``New instanton effects in string theory,'' {\em
  JHEP} {\bf 02} (2006) 060,
\href{http://www.arXiv.org/abs/hep-th/0512039}{{\tt hep-th/0512039}}.
%%CITATION = HEP-TH 0512039;%%.

\bibitem{Gutowski:1999tu}
J.~Gutowski, G.~Papadopoulos, and P.~K. Townsend, ``Supersymmetry and
  generalized calibrations,'' {\em Phys. Rev.} {\bf D60} (1999) 106006,
\href{http://www.arXiv.org/abs/hep-th/9905156}{{\tt hep-th/9905156}}.
%%CITATION = HEP-TH/9905156;%%.

\bibitem{Gmeiner:2006ni}
F.~Gmeiner and F.~Witt, ``Calibrated cycles and T-duality,''
\href{http://www.arXiv.org/abs/math.dg/0605710}{{\tt math.dg/0605710}}.
%%CITATION = MATH.DG/0605710;%%.

\end{thebibliography}\endgroup
\bibliographystyle{utphys} 
%\bibliographystyle{plain} 

%%%%%%%%%%%%%%%%%%%%%%%%%%%%%%%%%%%%%%%%%%%%%%% 
%%%%%%%%%%%%%%%%%%%%%%%%%%%%%%%%%%%%%%%%%%%%%%% 
%%%%%%%%%%%%%%%%%%%%%%%%%%%%%%%%%%%%%%%%%%%%%%% 
%%%%%%%%%%%%%%%%%%%%%%%%%%%%%%%%%%%%%%%%%%%%%%% 
%%%%%%%%%%%%%%%%%%%%%%%%%%%%%%%%%%%%%%%%%%%%%%% 
%%%%%%%%%%%%%%%%%%%%%%%%%%%%%%%%%%%%%%%%%%%%%%% 
%%%%%%%%%%%%%%%%%%%%%%%%%%%%%%%%%%%%%%%%%%%%%%% 
%%%%%%%%%%%%%%%%%%%%%%%%%%%%%%%%%%%%%%%%%%%%%%% 

\end{document}